\documentclass[aps,prx,twocolumn,floats,showpacs,superscriptaddress,nofootinbib]{revtex4-1}
\usepackage{graphicx,epsfig}
\usepackage{times,bbm}
\usepackage{graphics,dcolumn,bm,float}
\usepackage{amssymb,amsmath,rotate,color,amsfonts}
\usepackage[title,titletoc,toc]{appendix}
\usepackage{mathtools}
\usepackage{booktabs}
\usepackage{tcolorbox}
\usepackage[pagebackref=false,colorlinks,linkcolor=magenta,citecolor=blue,urlcolor=magenta]{hyperref}

\usepackage{wrapfig}
\usepackage{lipsum}
\usepackage{mwe}
\usepackage[mathlines]{lineno}
\usepackage[mathscr]{euscript}
\usepackage{hyperref}
\usepackage{breqn}
\usepackage{bbold}
\usepackage{pgfplots}
\usepackage{float}
\usepackage{tkz-euclide}
\usepackage{braket}
\usepackage{physics}
\usepackage[caption=false]{subfig}
\usepackage[export]{adjustbox}
\usepackage{tikz}
\usepackage{slashed}
\usepackage{bm}
\usepackage{multirow}
\usetikzlibrary{through,calc}
\usetikzlibrary{positioning}

\newcommand{\be}{\begin{equation}}
\newcommand{\ee}{\end{equation}}

\newcommand{\bea}{\begin{eqnarray}}
\newcommand{\eea}{\end{eqnarray}}
\newcommand{\nn}{\nonumber}

\newcommand{\bs}{\boldsymbol}
\newcommand{\bmt}{\left[\begin{matrix}}
\newcommand{\emt}{\end{matrix}\right]}

\begin{document}
\preprint{}
\title{Kinetic Theory of {\it Tilted} Dirac Cone Materials}

\author{A. Moradpouri}
\affiliation{Department of Physics$,$ Sharif University of  Technology$,$ Azadi Ave$,$ P.Code 1458889694$,$ Tehran$,$ Iran}

\author{S.A. Jafari}
\affiliation{Department of Physics$,$ Sharif University of  Technology$,$ Azadi Ave$,$ P.Code 1458889694$,$ Tehran$,$ Iran}

\author{Mahdi Torabian}
\email{mahdi.torabian@sharif.ir}
\affiliation{Department of Physics$,$ Sharif University of  Technology$,$ Azadi Ave$,$ P.Code 1458889694$,$ Tehran$,$ Iran}

\date{\today}
\begin{abstract}
We formulate the Boltzmann kinetic equations for interacting electrons with tilted Dirac cones in two space dimensions characterized by a tilt parameter 
$0\le\zeta<1$. By solving the linearized Boltzmann equation, we find that the broadening of the Drude pole is enhanced by $\kappa(\zeta)\times(1-\zeta^2)^{-1/2}$,
where the $\kappa$ is interaction-induced enhancement factor. The intensity of the Drude pole is also anisotropically enhanced by $(1-\zeta^2)^{-1}$. 
The ubiquitous "redshift" factors $(1-\zeta^2)^{1/2}$ can be regarded as a manifestation of an underlying spacetime structure in such solids. 
The additional broadening $\kappa$ that arises from the interactions can not be obtained from a simple coordinate change and
are more pronounced for electrons in a $\zeta$-deformed Minkowski spacetime of tilted Dirac fermions. 
\end{abstract}



\maketitle
\narrowtext

\section{Introduction}
Boltzmann equations are powerful method to study the transport and optical properties of electrons/hole plasmas~\cite{Jauho}.
Within this method one can address the coupling of charge carriers to impurities, phonons and their interactions with themselves~\cite{ziman2001electrons,Girvin2019}. Within this method, the relativisitc fermions can also be addressed~\cite{Cercignani2002}, 
enabling the study of transport in Dirac materials examplified in two dimensions by graphene~\cite{SarmaReview}. 
Furthermore, Berry phase effects can be easily accomodate within this method~\cite{Niu2010}. 
Application of kinetic theory to Dirac/Weyl materials in three space that host 
chiral fermions\footnote{note that in two space dimensions there is no $\gamma^5$ matrix and hence no chirality 
in the sese of left/right-handedness can be defined}~can powerfully capture the 
chiral anomaly and its consequences~\cite{Stephanov2012,SonSpivak2013}.

When the chemical potential in a Dirac material is zero, and it is defectless~\cite{kashuba}, 
it serves as a realization of a quantum critical 
system~\cite{SachdevBook,sachdev08}, where having set the chemical potential $\mu=0$, the physics is entirely determined by 
the ration of $k_BT$ and $\hbar\omega$, the frequency at which the system is being probed. In the undoped Dirac materials,
the kinetic theory can be used to study the ration $\eta/s$ of the viscosity and entropy density that suggests 
graphene as a perfect fluid~\cite{muller2009graphene,lucas2018hydrodynamics}, which has been indeed evidenced in 
experiments~\cite{Crossno2016,Bandurin2016,Bandurin2018}. 

A variant of Dirac materials are materials with tilted Dirac cone in their spectrum that can occure in 
various classes of materials from 8Pmmn-borohpene~\cite{Zhou2014,Lopez2016} to MoS$_2$ family~\cite{MoS2Tilted}.
Original observation of tilted Dirac cone dispersion was in the organig compound (BEDT-TTF)$_2$I$_3$~\cite{Kajita2014Review,Katayama2006,Suzumura2012}. 
The Hamiltonian of these systems (in 2+1 dimensional spacetime) is given by
\be
   H=v_F\bs\sigma.\bs p+v_F\bs\zeta.\bs p
   \label{tiltHamiltonian.eqn}
\ee
where the second term is characterized by a vector-looking dimensionless parameter $\bs \zeta$ that quantifies the 
tilting of the Dirac cone and we limit our focus to $|\bs \zeta|<1$ situation. The above innocent looking energy dispersion
has far reaching consequnces. To see how, let us start with the Hamiltonian of an upright Dirac cone given 
by $H'=v_F\bs\sigma.\bs p$. Now imagine the "coordinate transformations" $t'=t$, ${\bs x'}=\bs x+ v_F\bs \zeta t$ 
known as Gallilean transformation, 
which implies $\partial_{\bs x'}=\partial_{\bs x}$ and $\partial_{t'}=\partial_t+v_F\bs\zeta.\bs\partial_{x}$.
The last equation readily suggests $H'=H+v_F\bs\zeta.\bs p$ which is nothing but Eq.~\eqref{tiltHamiltonian.eqn}. 
Affecting the above coordinate transformation on the Minkiwski metric $\eta_{ab}=\rm{diag}(-1,1,1)$ 
or $d{s}^2=-v_F^2d{t'}^2+(d{\bs x'})^2$ of the Dirac materials gives,
\be
   ds^2=-v_F^2dt^2+(d\bs x-v_F\bs \zeta dt)^2.
   \label{metricds.eqn}
\ee
This is indeed the emergent metric $g_{\mu\nu}$ that consistently describes the tilted Dirac/Weyl cone 
materials~\cite{Volovik2016,Tohid2019Spacetime,Ojanen2019,SaharCovariance,Jafari2019}.
The matrix representation of the metric~\eqref{metricds.eqn} can be diagonalized by an orthogonal matrix $e^a_\mu$ as
\be
    g_{\mu\nu}=e_{\mu}^a e_{\nu}^b \eta_{ab}.
    \label{vier.eqn}
\ee
The above equation in fact shows that the Dirac fermions in tilted Dirac cone materials are equpped with 
"frame fields" $e_\mu^a$~\cite{RyderGR,SchutzGR,weinbergbook}.
This indicates that the tilt is in fact a proxy for the frame fields, hence a 
deep gravitational analogy is hidden in tilted Dirac cone materials. 

Therefore it is timely to formulate a kinetic theory for the fermions in tilted Dirac materials~\cite{KineticCurved}.
As in earlier works~\cite{sachdev08,muller2009graphene} we consider the undoped case $\mu=0$, where like any
other quantum critical system the relaxation time of electron-electron interaction
(Coulomb interaction) is controlled by the $k_BT$ and the frequency $\omega$ at which the system is probed as
$\tau^{-1}_{\rm ee}\sim\alpha^2\frac{k_BT}{\hbar}$~\cite{lucas2018hydrodynamics,SachdevBook}. 
On the other hand, the relaxation time for a (dilute) density of charged impurities is given by~\cite{muller2008quantum}
$\tau^{-1}_{\rm imp}\sim\frac{1}{\hbar}\frac{(\frac{Ze^2}{\epsilon_r})^2\rho_{\rm imp}}{{\rm max}[k_BT,\mu]}$,
where $Ze$ is the charge of an random impurity and $\rho_{\rm imp}$ is the 
average spatial density of impurities. We see that in the high temperature limit $k_BT\gg\mu$ (which specifies the 
appropriate regime for Dirac fluid and applies to our case) and for a dilute density of impurities, the Coulomb interaction 
dominates the relaxation mechanism ($\tau^{-1}_{\rm ee}\gg\tau^{-1}_{\rm imp}$). 
So in this limit, we may ignore the effect of disorder under some mild assumptions (dilute density, no localization).
Furthermore at $\mu=0$ limit of interest to us, at zero temperature there will be no screening effect arising from the Dirac electrons.
But at elevated temperature there can be thermally excited population of electrons and holes that may lead to classical (Debye) screening. 
Since our kinetic theory formulation is second order in Coulomb interaction ($\alpha$), consideration of screening in this case will correspond
to higher order effects in $\alpha$. Therefore to be consistent at order $\alpha^2$, we ignore the screening effects even at elevated temperatures.
So we can focus on the sole role of Coulomb interactions. 
It is important to note that the Coulomb interactions can not be covariantly written in terms of the above Gallilean transformations.
This is because despite an emergent spacetime structure of the form~\eqref{metricds.eqn}, the photons mediating the Coulomb 
interaction mostly pass through the vacuum. Therefore, although some single particle properties like density of states
at non-zero chemical potential can be obtained from the Jacobian of the above transformation, the many-body properties
in the presence of Coulomb forces do not conform to this logic. 

Investigation of the $\mu=0$ tilted Dirac cone systems at non-zero temperature will allow us to ignore the impurities,
thereby to study the effects that purely arise from the spacetime structure~\eqref{metricds.eqn} combined with Coulomb
interactions. The general outcome is that, the presence of tilt enhances
the many-body effects as manifested in the broadening of the Drude peak, in agreement with earlier study using independent method~\cite{jalali2021tilt}. 
In section~\ref{formulation.sec} we introduce the Hamiltonian and notations.
In section~\ref{transport.sec} we consider the collisionless limit of the Boltzmann equations for Dirac fermions 
in the background metric~\eqref{metricds.eqn}. In section~\ref{viscous.sec} we consider the effect of Coulomb interactions
in such a spacetime background. We end the paper by discussions and outlook in section~\ref{discuss.sec}. 

\section{Hamiltonian and notations}
{\label{formulation.sec}}
We study the kinetic theory of electrons in tilted Dirac cone materials with Coulomb interactions. The total Hamiltonian consists of two terms: the free part $H_0$ and the interaction term $H_{\rm Coulomb}$. The free Hamiltonian of tilted Dirac electrons is given by
\begin{equation}
	H_0=\psi^{\dagger a}
	\begin{pmatrix}
		\vec p\cdot \vec\zeta & (p_x-ip_y) \\
		(p_x+ip_y) & \vec p\cdot \vec\zeta
	\end{pmatrix}
	\psi^a,
\end{equation}
where $a$ accouts for spin-valley index. The interaction Hamiltonian is assumed to be the Coulomb electron-electron interaction given by
\begin{equation}
	\begin{aligned}
			H_{\rm Coulomb}&=
			\frac{1}{2}\int \frac{{\rm d}^2\vec p_1}{(2\pi)^2}\frac{{\rm d}^2\vec p_2}{(2\pi)^2}\frac{{\rm d}^2\vec q}{(2\pi)^2}\\
			&\times\psi^{\dagger a}_{\vec p_1-\vec q}\psi^a_{\vec p_1}V(q)\psi^{\dagger b}_{\vec p_2+\vec q}\psi^b_{\vec p_2},
			\end{aligned}
\end{equation}
where $V(q)$ is the static Coulomb potential
\begin{equation}
	V(q)=\frac{2\pi e^2}{\epsilon|\vec q|},
\end{equation} 
and $\epsilon$ is the dielectric constant of the medium. 
The free Hamiltonian can be diagonalized through a unitary transformation $H_0^{\rm diag}=U^{\dagger}H_0U$. The transformation matrices are 
\begin{equation}
	\begin{aligned}
	U&=\frac{1}{\sqrt 2}
	\begin{pmatrix}
		1 & 1\\
		 e^{i\phi_p}&  -e^{i\phi_p}\\
	\end{pmatrix},
\end{aligned}
\end{equation} 
where $\phi_p$ is the polar angle of vector $\vec p$. In the diagonal basis, the electron field is transformed into
\be \psi_{\vec p}=\begin{pmatrix}
	\psi_{1,\vec p}\\
	\psi_{2,\vec p}\\
\end{pmatrix} \rightarrow \psi^{\rm diag}_{\vec p}= U^{\dagger}\psi_{\vec p}=\begin{pmatrix}
	c_{+,\vec p} \\
	c_{-,\vec p}\\
\end{pmatrix},
\ee
where 
\begin{equation}
	\begin{aligned}
	&&c_{+,\vec p}=\frac{1}{\sqrt 2}(\psi_{1,\vec p}+\psi_{2,\vec p}),\\
	&&c_{-,\vec p}=\frac{P}{|\vec p|}\frac{1}{\sqrt2}(\psi_{1,\vec p}-\psi_{2,\vec p}),
	\end{aligned}
\end{equation} 
and $P=p_x+ip_y$.
Then, the diagonal free Hamiltonian is given by
\begin{equation}
\begin{aligned}	
&&H^{\rm diag}_0=\sum_{a,s}\int\frac{{\rm d}^2\vec p}{(2\pi)^2}E_{s,\vec p}c_{a,s,\vec p}^{\dagger}c_{a,s,\vec p},
\end{aligned}	
\end{equation}
where 
\be E_{s,\vec p}=sv_F|\vec p| +v_F\vec p\cdot\vec\zeta,\ee
and $s=\pm$ is the band index denoting upper ($+$) and lower ($-$) branches of the tilted Dirac cone, 
while as mentioned,  $a$ is the spin-valley index. 

The interaction Hamiltonian in diagonal basis $c_{a,s}$ becomes
\begin{equation}
	\begin{aligned}	
		H_{\rm Coulomb}=&\!\!\!\!\!\sum_{s_1,s_2,s_3,s_4}\!\int \frac{{\rm d}^2\vec p_1}{(2\pi)^2}\frac{{\rm d}^2\vec p_2}{(2\pi)^2}\frac{{\rm d}^2\vec q}{(2\pi)^2}T_{s_1s_2s_3s_4}(k_1,k_2,q)\\ 
		&\quad\times c_{b,s_4,\vec p_1+\vec q}^{\dagger }c_{a,s_3,\vec p_2-\vec q}^{\dagger }V(q)c_{a,s_2,\vec p_2}c_{b,s_1,\vec p_1}\ ,
	\end{aligned}	
\end{equation}
with
\begin{equation}
	\begin{aligned}
		T_{s_1s_2s_3s_4}(p_1,p_2,q)=\frac{1}{8}V(q)&\times\Big(1+s_1s_4\frac{(P^*_1+Q^*)P_1}{| \vec p_1+\vec q||\vec p_1|}\Big)\\
		&\times\Big(1+s_2s_3\frac{(P^*_2-Q^*)P_2}{|\vec p_2-\vec q||\vec p_2|}\Big),
	\end{aligned}
\end{equation}
where $s_1,s_2$ and $s_3,s_4$ are the band indices of incoming and outgoing particles, respectively
and $Q=q_x+iq_y$. Moreover, electric current  $J^i=-ev_F\bar\psi\gamma^i\psi$ in the diagonal basis is
\begin{equation}
\begin{aligned}
\vec J=e\nu_F\sum_s\int \frac{{\rm d}^2\vec p}{(2\pi)^2} \Big[&\Big(\vec\zeta+s\frac{\vec p}{|\vec p|}\Big)c^{\dagger}_{s,\vec p}c_{s,\vec p} \\ &+ i\frac{\hat z \times \vec p}{|\vec p|}sc^{\dagger}_{s,\vec p}c_{-s,\vec p}\Big],
\label{currentequation}
\end{aligned}
\end{equation}
where $\hat z$ is the unit vector transverse to the $xy$ plane of the material hosting the tilted Dirac fermions. 

\section{Collisionless Limit }
{\label{transport.sec}}
To set the stage, 
in this section by turning off the Coulomb interactions, we study collisionless limit of the transport equation for the intraband and interband transition processes. To calculate electric transport coefficient, we apply an electric field. The interaction Hamiltonian between the applied electric field and 
the tilted Dirac fermions is~\cite{katsnelson2007} 
\begin{equation}
	\begin{aligned}
&		H_{\rm int}=-e\vec E\cdot \big(i\psi^{\rm diag}_{\vec p}\vec\partial \psi^{\rm diag}_{\vec p}\big) \\
		&=-e\vec E\cdot \sum_{a,s}\int \frac{{\rm d}^2\vec p}{(2\pi)^2}\Big(ic_{a,s,\vec p}^{\dagger}\vec\partial c_{a,s,\vec p}+\frac{1}{2}c_{a,s\vec p}^{\dagger}c_{a,-s,\vec p}\vec\partial \phi_{\vec p}\Big).
	\end{aligned}
\end{equation}
 
For later applications, we define the band diagonal (off-diagonal) distribution functions $f$ ($g$) as follows
\begin{equation}
	\begin{aligned}
	&f_s(\vec p,t)=\langle c^{\dagger}_{a,s,\vec p}c_{a,s,\vec p}\rangle,\\
&g_s(\vec p,t)=\langle c^{\dagger}_{a,s,\vec p}c_{a,-s,\vec p}\rangle.
	\end{aligned}
\end{equation}
There is no sum over spin-valley index $a$. We assume that these distribution functions are the same for all spin and valley indices.
We use the Liouville equation~\cite{pathria}
\begin{equation}
	\begin{aligned}
	\frac{{\rm d}{\cal O}}{{\rm d}t}=\frac{1}{i\hbar}[{\cal O},H^{\rm diag}_0+H_{\rm int}],
	\end{aligned}
\end{equation}
to find the (generalized) Boltzmann equation for the distribution functions that correspond to intraband ($f$) and interband ($g$) transition as
\begin{equation}
\begin{aligned}	
&\frac{\partial f_s(\vec p,t)}{\partial t}+e\vec E\cdot\vec\partial_p f_s(\vec p,t)\\
& -i\frac{e\vec E\cdot\vec\partial_p\phi_p}{2|\vec p|}\Big(g_s(\vec p,t)-g_{-s}(\vec p,t)\Big)=0,
\end{aligned}
\label{f-equation}	
\end{equation}
and 
\bea
&&\frac{\partial g_s(\vec p,t)}{\partial t}-i\big(E_{s,\vec p}-E_{-s,\vec p}\big)g_s(\vec p,t)+e\vec E\cdot \vec\partial_p g_s(\vec p,t)\nn\\
&&-i\frac{e\vec E\cdot \vec\partial_p\phi_p}{2|\vec p|}\Big(f_s(\vec p,t)-f_{-s}(\vec p,t)\Big)=0.
\label{g-equation}
\eea
Perturbative solution can be achieved by expanding the distribution functions as
\bea
&&f_s(p,t)=f^0_s(p,t)+f^1_s(p,t)+\cdots,\\
&&g_s(p,t)=g^0_s(p,t)+g^1_s(p,t)+\cdots.
\eea
At the leading order $f_s$ is the Fermi-Dirac function 
\be f^0_s(E_{s,\vec p})=\frac{1}{1+e^{\beta \gamma E_{s,\vec p}}},\ee
where $\gamma = {(1-\zeta^2)}^{-1/2}$. Moreover, as can be seen from~\eqref{g-equation}, at the leading order, $g_s(\vec p,s)$ is zero at equilibrium, namely $g_s^0(\vec p,t)=0$, so at the first order the last term in \eqref{f-equation} is vanishing. Thus, the distribution function $f$ upto the first order is given by
\begin{equation}
	f_{s}(\vec p,\omega)=2\pi\delta(\omega)f^{0}(E_{s})-e\vec E\cdot \vec\partial_p f^{0}(E_{s})g(\vec p,\omega).
	\label{ansatz}
\end{equation}

 To compute the interband transition from Eq.~\eqref{currentequation} we need to find imaginary part of the distribution function $g_s(\vec p,t)$,
 \be g_s(\vec p,t)=g'_{s}(\vec p,t)+i g''_{s}(\vec p,t).\ee
To leading order, solutions to Eqs.~\eqref{f-equation} and ~\eqref{g-equation} are
 \begin{equation}
 	\begin{aligned}	
 	&g''_{+}(\vec p,\omega)=\frac{ie\omega\big(f_{+}^0(\vec p)-f_{-}^0(\vec p)\big)E_x(\omega)}{2|\vec p|(4v^2_F|\vec  p|^2-\omega^2)},\\
 	&g(\vec p,\omega)=\frac{1}{-i\omega+\delta},
 	\end{aligned}
 \end{equation}
 where we have Fourier transformed from time to frequency with $\delta$ being an infinitesimal positive number.
The electric field is applied along the $x$ axis. 

Finally, the electric transport coefficients for intraband transition are computed as follows:
\bea	
\sigma_{\rm intra}^{xx}&=&\frac{e^2}{h}\frac{2Nk_BT (1-\zeta^2)^{\frac{1}{2}}\ln2}{(-i\hbar\omega+\delta)\zeta^2}\cr& \times&\Big[-\frac{\zeta^2\zeta_x^2-\zeta_y^2}{(1-\zeta^2)^{\frac{3}{2}}}\!+\!\frac{1}{\zeta^2}\Big(1+\frac{2\zeta^2-1}{(1-\zeta^2)^{\frac{3}{2}}}\Big)(\zeta_x^2-\zeta_y^2)\Big],\quad\ \\
\sigma_{\rm intra}^{xy}&=&\frac{e^2}{h}\frac{2Nk_BT (1-\zeta^2)^{\frac{1}{2}}\ln2}{(-i\hbar\omega+\delta)\zeta^2}\cr& \times&\Big[\frac{2}{\zeta^2}\Big(1+\frac{2\zeta^2-1}{(1-\zeta^2)^{\frac{3}{2}}}\Big)-\frac{1+\zeta^2}{(1-\zeta^2)^{\frac{3}{2}}}\Big]\zeta_x\zeta_y,
\eea
while for the interband transitions we have
\bea	\sigma^{xx}_{\rm inter}&=&\frac{e^2N}{8h}M_1(a,\zeta,\alpha),\\
	\sigma^{xy}_{\rm inter}&=&-\frac{e^2N}{8h}M_2(a,\zeta,\alpha),
\eea
where
\bea		
M_1[\bar\omega,\zeta,\alpha]&&=\int d\phi_p\Big[\frac{1}{1+e^{\frac{\hbar\omega}{2k_BT}\gamma(-1+\zeta\cos(\phi_p-\alpha))}}\cr
		&&\qquad-\frac{1}{1+e^{\frac{\hbar\omega}{2k_BT}\gamma(1+\zeta \cos(\phi_p-\alpha))}}\Big]\sin(\phi_p)^2,\ \ \ \ \  \\
M_2[\bar\omega,\zeta,\alpha]&&=\int d\phi_p\Big[\frac{1}{1+e^{\frac{\hbar\omega}{2k_BT}\gamma(-1+\zeta\cos(\phi_p-\alpha))}}\cr
		-&&\frac{1}{1+e^{\frac{\hbar\omega}{2k_BT}\gamma(1+\zeta \cos(\phi_p-\alpha))}}\Big]\sin(\phi_p)\cos(\phi_p).\ \ \ \ 
\eea
Here $\bar\omega=\frac{\hbar\omega}{k_BT}$ and $\alpha$ is the angle of $\vec\zeta$ with respect to the $x$-axis. In the low frequency limit $\omega\to 0$, we have 
\bea		
M_1[a,\zeta,\alpha]=M_2[a,\zeta,\alpha]\to 0,
\label{interbandcontributions}
\eea
and interband contributions go to zero as they must.

\section{Transport in presence of Coulomb interaction}
{\label{viscous.sec}}
After the warmup in section~\ref{transport.sec} with non-interacting tilted Dirac cone fermions, we are now ready to turn on the Coulomb interactions
and investigate the fate of Coulomb interaction for tilted Dirac-cone fermions in order to find whether the Coulomb interactions play more important 
role in tilted Dirac fermions or not. In fact earlier investigation within the Fermi liquid theory shows that the Coulomb interactions play more 
important role for tilted Dirac fermions rather than the non-tilted Dirac fermions~\cite{jalali2021tilt}. Therefore it is useful to 
study the role of Coulomb interactions within the kinetic theory approach. In particular we will focus on the $\mu=0$ and high enough 
temperatures to ensure that the disorder does not play a significant role, and the Coulomb forces will be the major players~\cite{sachdev08,muller2009graphene}.
As can be seen in Eq.~\eqref{interbandcontributions}, 
in the low frequency limit interband transition can be ignored. Here we restrict ourselves to this regime. The Boltzmann equation include scattering terms induced by Coulomb interaction that are given by
\begin{widetext}
\begin{equation}
\begin{aligned}	
&\frac{\partial f_s(\vec p,t)}{\partial t}+e\vec E\cdot\vec\partial_pf_s(\vec p,t)
=-\frac{2\pi}{v_F}\int\frac{{\rm d}^2\vec p_1}{(2\pi)^2}\frac{{\rm d}^2\vec q}{(2\pi)^2}\\& \bigg(\delta(E_{s,\vec p}+E_{-s,\vec p_1}-E_{s,\vec p+\vec q}-E_{-s,\vec p_1-\vec q})R_1(\vec p,\vec p_1,\vec q)\times\\
&\times \Big[f_s(\vec p,t)f_{-s}(\vec p_1,t)\Big(1-f_s(\vec p+\vec q,t)\Big)\Big(1-f_{-s}(\vec p_1-\vec q,t)\Big)-f_s(\vec p+\vec q,t)f_{-s}(\vec p_1-\vec q,t)\Big(1-f_s(\vec p,t)\Big)\Big(1-f_{-s}(\vec p_1,t)\Big)\Big]\\
&\!\! -\delta(E_{s,\vec p}+E_{s,\vec p_1}-E_{s,\vec p+\vec q}-E_{s,\vec p_1-\vec q}))R_2(\vec p,\vec p_1,\vec q)\times\\
&\times \Big[f_s(\vec p,t)f_s(\vec p_1,t)\Big(1-f_s(\vec p+\vec q,t)\Big)\Big(1-f_s(\vec p_1-\vec q,t)\Big)-f_s(\vec p+\vec q,t)f_s(\vec p_1-\vec q,t)\Big(1-f_s(\vec p,t)\Big)\Big(1-f_s(\vec p_1,t)\Big)\Big]\bigg),\\	
\end{aligned}
\label{Boltzmann.eqn}
\end{equation}  
\end{widetext}
where $R_1(\vec p,\vec p_1,\vec q)$ and $R_2(\vec p,\vec p_1,\vec q)$ are scattering amplitude functions for particles of opposite (i.e. electron-hole) and identical
(i.e. electron-electron or hole-hole) charge, respectively which are given in the appendix~\ref{AppendixA}.
We linearize the distribution function $f_s(\vec p,\omega)$ in Boltzmann equation~\eqref{Boltzmann.eqn} and use the ansatz~\eqref{ansatz} 
to obtain 
\begin{widetext}
\begin{equation}
\begin{aligned}	
&\frac{\beta'v_F(-i\omega g(\vec p,\omega)-1)}{(1+e^{\beta'E_{s,\vec p}})(1+e^{-\beta'E_{s,\vec p}})}\Big(s\frac{\vec p}{p}+\vec\zeta\Big)=-2\pi\int\frac{{\rm d}^2\vec p_1}{(2\pi)^2}\frac{{\rm d}^2\vec q}{(2\pi)^2}\\&
\bigg[\frac{\delta(E_{s,\vec p}+E_{-s,\vec p_1}-E_{s,\vec p+\vec q}-E_{-s,\vec p_1-\vec q})R_1(\vec p,\vec p_1,\vec q)}{(1+e^{-\beta'E_{s,\vec p}})(1+e^{-\beta'E_{-s,\vec p_1}})(1+e^{\beta'E_{s,\vec p+\vec q}})(1+e^{\beta'E_{-s,\vec p_1-\vec q}})}\times
\\&\qquad\times\Big(\frac{\partial E_{s,\vec p}}{\partial\vec p}g(\vec p,\omega)+\frac{\partial E_{-s,\vec p_1}}{\partial\vec p_1}g(\vec p_1,\omega)-\frac{\partial E_{s,\vec p+\vec q}}{\partial(\vec p+\vec q)}g(\vec p+\vec q,\omega)
-\frac{\partial E_{-s,\vec p_1-\vec q}}{\partial(\vec p_1-\vec q)}g(\vec p_1-\vec q,\omega)\Big)\\
&+\frac{\delta(E_{s,\vec p}+E_{s,\vec p_1}-E_{s,\vec p+\vec q}-E_{s,\vec p_1-\vec q}))R_2(\vec p,\vec p_1,\vec q)}{(1+e^{-\beta'E_{s,\vec p}})(1+e^{-\beta'E_{s,\vec p_1}}(1+e^{\beta'E_{s,\vec p+\vec q}})(1+e^{\beta'E_{s,\vec p_1-\vec q}})}\times \\
&\quad \times\Big(\frac{\partial E_{s,\vec p}}{\partial\vec p}g(\vec p,\omega)+\frac{\partial E_{s, \vec p_1}}{\partial\vec p_1}g(\vec p_1,\omega)-\frac{\partial E_{s,\vec p+\vec q}}{\partial(\vec p+\vec q)}g(\vec p+\vec q,\omega)-\frac{\partial E_{s,\vec p_1-\vec q}}{\partial(\vec p_1-\vec q)}g(\vec p_1-\vec q,\omega)\Big)\bigg],
\end{aligned}
\label{LBoltzmann.eqn}
\end{equation} 
\end{widetext}
where $\beta'$ is defined as $\beta'=\beta\gamma$. Finding an analytic solution to the above integro-differential equation even in the linearized limit
is formidable task. Therefore, we numerically solve the Boltzmann equation as discussed in the following. 

There are points in the momentum space where the Coulomb interaction is singular and need to be handled with special care. These points are given by 
$\vec q\to 0$ and $\vec q\to\vec p_1-\vec p$ where the Coulomb potential $V(\vec q)$ and $V(\vec p_1-\vec p-\vec q)$ diverge as inverse square of the distance
to $0$ and $\vec p_1-\vec p$, respectively. This is because
the Coulomb matrix element appears as the square of the $T$-matrices defining $R_1(\vec p,\vec p_1,\vec q)$ and $R_2(\vec p,\vec p_1,\vec q)$ (see appendix~\ref{AppendixA}).
However, if we expand functions in~\eqref{LBoltzmann.eqn} around $\vec q=0$ and $\vec q=\vec p_1-\vec p$, these points are zeros of those functions. Given the angular integration, the first order term does not contribute to the integral, leaving behind a second order term in the numerator. Thus the integral is regular around these points and can be numerically integrated. The other source of divergences, are the anti-collinear and collinear scattering processes, which are the characteristic feature of linear energy dispersion in 2D materials~\cite{sachdev97}.

The Boltzmann equation for time reversal invariant interactions can be derived from the variational principles~\cite{ziman2001electrons,Arnold2000} which is explored in more detail in appendix~\ref{AppendixB}. Generally, we can cast the linearized Boltzmann equation into the operator formulation as~\cite{muller2009graphene}
\begin{equation}
	\begin{aligned}	
	\ket{\phi}=C\ket{g},
	\end{aligned}
\end{equation}
where $|g\rangle$ defines the perturbed distribution function and $C$ defines the collision processes and $|\phi\rangle$ defines external perturbations such 
as the electric field and/or temperature gradients~\cite{muller2009graphene}. 
To find the perturbed state $\ket{g}$, we need to find inverse of collision operator $C^{-1}$. Following~\cite{sachdev08}, if $\lambda_n$ are the eigenvalues of collision operator $C$,  then $C^{-1}$ has the followin spectral representation:
\begin{equation}
		C^{-1}=\sum_n\frac{1}{\lambda_n}\ket{n}\bra{n},
		\label{spectral.eqn}
\end{equation}
where $\ket{n}$ is the $n$'th eigenvector of collision operator $C$. It is clear that
the perturbed state $|g\rangle$ are dominated by the smallest eigenvalue of the collision operator $C$.  
 
In the case of Coulomb interaction, the right hand side of Eq.~\eqref{LBoltzmann.eqn} can be viewed as a linear operator acting on 
$\frac{\partial E_{s,\vec p}}{\partial \vec p}g(\vec p,\omega)$. So, one can define an inner product with respect to which the above operator is Hermitian as follows:
\begin{equation}
	\bra{g_1}\ket{g_2}=\sum_s\int \frac{d^2p}{(2\pi)^2}g_{1s}(\vec p)g_{2s}(\vec p).	
\end{equation}  
Based on this, we introduce a functional whose stationary solution are the Boltzmann Eq.~\eqref{LBoltzmann.eqn} as follows:
\begin{widetext}
	\begin{equation}
		\begin{aligned}	
			&{\cal Q}(g)=\frac{2\pi}{8}\int\frac{{\rm d}^2p}{(2\pi)^2}\frac{{\rm d}^2p_1}{(2\pi)^2}\frac{{\rm d}^2q}{(2\pi)^2}
			\bigg[\frac{\delta(E_{s,\vec p}+E_{-s,\vec p_1}-E_{s,\vec p+\vec q}-E_{-s,\vec p_1-\vec q})R_1(\vec p,\vec p_1,\vec q)}{(1+e^{-\beta'E_{s,\vec p}})(1+e^{-\beta'E_{-s,\vec p_1}})(1+e^{\beta'E_{s,\vec p+\vec q}})(1+e^{\beta'E_{-s,\vec p_1-\vec q}})}\times
			\\&\qquad\qquad\qquad\qquad\qquad\qquad\times\frac{1}{v_F^2}\Big(\frac{\partial E_{s,\vec p}}{\partial\vec p}\hat g(\vec p,\omega)+\frac{\partial E_{-s,\vec p_1}}{\partial\vec p_1}\hat g(\vec p_1,\omega)-\frac{\partial E_{s,\vec p+\vec q}}{\partial(\vec p+\vec q)}\hat g(\vec p+\vec q,\omega)
			-\frac{\partial E_{-s,\vec p_1-\vec q}}{\partial(\vec p_1-\vec q)}\hat g(\vec p_1-\vec q,\omega)\Big)^2\\
			&\qquad\qquad\qquad\qquad\qquad\qquad\qquad+\frac{\delta(E_{s,\vec p}+E_{s,\vec p_1}-E_{s,\vec p+\vec q}-E_{s,\vec p_1-\vec q}))R_2(\vec p,\vec p_1,\vec q)}{(1+e^{-\beta'E_{s,\vec p}})(1+e^{-\beta'E_{s,\vec p_1}})(1+e^{\beta'E_{s,\vec p+\vec q}})(1+e^{\beta'E_{s,\vec p_1-\vec q}})}\times\\
			&\qquad\qquad\qquad\qquad\qquad\qquad\times\frac{1}{v_F^2}\Big((\frac{\partial E_{s,\vec p}}{\partial\vec p}\hat g(\vec p,\omega)+\frac{\partial E_{s,\vec p_1}}{\partial\vec p_1}\hat g(\vec p_1,\omega)-\frac{\partial E_{s,\vec p+\vec q}}{\partial(\vec p+\vec q)}\hat g(\vec p+\vec q,\omega)-\frac{\partial E_{s,\vec p_1-\vec q}}{\partial(\vec p_1-\vec q)}\hat g(\vec p_1-\vec q,\omega)\Big)^2\bigg]\\
			&\qquad+\int\frac{{\rm d}^2p}{(2\pi)^2}\frac{\hat g(\vec p,\omega)(-\frac{i\omega \hat g(\vec p,\omega)}{2}-\beta'v_F)}{(1+e^{\beta'E_{s,\vec p}})(1+e^{-\beta'E_{s,\vec p}})}\Big(1+s\frac{\vec p\cdot \vec\zeta}{p}\Big),
		\end{aligned}
	\end{equation} 
	\label{variational}
\end{widetext}
where $\hat g(p,\omega)=\beta'v_F g(p,\omega)$.

Due to momentum conservation, the delta function of energy conservation of tilted Dirac matterial is the same as graphene.
Delta function for energy conservation of same-charge particle scattering defines an ellipse for the set of points $\vec q=(q_x,q_y)$ where the sum of distances from two canonical points $\vec p_1$ and $-\vec p$ is constant, namely:
\be
|\, \vec p\,| +|\, \vec p_1|- |\,\vec p+\vec q\,|- |\, \vec p_1-\vec q\,|=0. 
\ee
We  parameterize the ellipse by elliptic coordinates $0\leq\mu<\infty$ and $ 0\leq\theta<2\pi$
in terms of which one can write~\cite{sachdev97}:
\begin{equation}
		Q=\frac{P_1-P}{2}+\frac{P_1+P}{4}(e^{\mu+i\theta}+e^{-\mu-i\theta}),	
	\label{ellipse}
\end{equation}
where $Q=q_x+iq_y$ and similarly for $P$ and $P_1$.
There is a constraint for delta function of same-charge particles as
\begin{equation}
		\cosh(\mu)=\frac{p+p_1}{|\,p+p_1|}.	
\end{equation}

For opposite-charge particles, the argument of the delta function of energy conservation defines a hyperbola for the set of points $\vec q=(q_x,q_y)$ where the differences of distances from two canonical points $\vec p_1$ and $-\vec p$ is constant, namely:
\bea
|\, \vec p\,| -|\,\vec p_1|-|\,\vec p+\vec q\,|+|\, \vec p_1-\vec q\,|&=&0.	
\eea
Depending on the relative absolute value of $p_1$ and $p$, there is a similar constraint  (e.g. for $p>p_1$)
\begin{equation}
	\begin{aligned}
		\cos(\theta)=\frac{p-p_1}{|\, p+p_1|}.	
	\end{aligned}
\end{equation}
So we need to choose $\theta$ range appropriately to be consistent with $p>p_1$ (or $p_1>p$).
For the same-charge particle, we can integrate over momentum in the elliptic coordinate $\mu$ 
\begin{equation}
	\begin{aligned}
	&\int\frac{{\rm d}^2\vec q}{(2\pi)^2}2\pi\delta(\vec p+\vec p_1+|\,\vec p+\vec q\,|-|\,\vec p_1-\vec q\,|)=\\
	&\qquad \int^{2\pi}_0\frac{{\rm d}\theta}{2\pi}\frac{|\,\vec  p+\vec p_1\,|}{4\sinh(\mu)}\Big(\cosh^2(\mu)-\cos^2(\theta)\Big).
	\end{aligned}
\end{equation}
In the collinear limit $\vec p=(p,0)$ and $\vec p_1=(p_1,p_{1\bot})$, we find
\begin{equation}
\begin{aligned}
\frac{|\, \vec p+\vec p_1|}{\sinh(\mu)}=(p+p_1)^2\sqrt{\frac{p_1}{p}}\frac{1}{p_{1\bot}}.	
\end{aligned}
\end{equation}
It is clear that integration over $p_{1\bot}$ is logarithmically divergent and the Coulomb interaction in the collinear limit can not cancel this singularity. 
This divergence is expected to be cutoff by higher order corrections to the self energy and will be proportional to structure constant $\alpha$ in the next order in the perturbation theory. The dominant contribution to the integral is in the range $\frac{T}{v_F}$ and $\frac{\alpha T}{v_F}$~\cite{Arnold2000}. Thus one estimates, 
 \begin{equation}
 	\begin{aligned}
 		\int {\rm d}p_{1\bot}\frac{1}{p_{1\bot}}\approx 2~\ln{\alpha}.
 	\end{aligned}
 \end{equation}
To study the collinear limit in more detail, we look at linearized Boltzmann equation~\eqref{LBoltzmann.eqn} for collision part of same-charge particles ~\cite{sachdev08,kashuba}. For the collinear limit, we choose the momentum $\vec p=(p,0)$ with $p>0$ and further we write $\vec p_1=(p_1,p_{1\bot})$, $\vec q=(q,q_{\bot})$ where $p_{1\bot}$ and $q_{\bot}$ are small. In the regime where $p>0$, $p+q>0$ and $p_1-q>0$, the argument of the energy conservation delta function can be simplified as:   
 \begin{equation}
	\begin{aligned}
	&|\, \vec p\,| +|\, \vec p_1|- |\,\vec p+\vec q\,|- |\, \vec p_1-\vec q\,|\\
	&\approx \frac{p^2_{1\bot}}{2p_1}-\frac{q^2_{\bot}}{2(p+q)}-\frac{(p_{1\bot}-q_{\bot})^2}{2(p_1-q)}=\\
	&-\frac{p_1+p}{2(p+q)(p_1-q)}(q_{\bot}-A_{+}p_{1\bot})(q_{\bot}-A_{-}p_{1\bot})
	\end{aligned}
\end{equation}
where $A_{\pm}$ are defined by:
 \begin{equation}
		A_{\pm}=\frac{p+q}{p+p_1}\pm\sqrt{\frac{p(p_1-q)(p+q)}{p_1(p+p_1)^2}}. 
\end{equation}
So the contribution arising from $R_2(\vec p,\vec p_1,q)$ to the Boltzmann equation becomes:
\begin{widetext}
\begin{equation}
	\begin{aligned}	
	\approx
		&\frac{8\alpha^2(\ln\alpha)\nu_F}{\pi}\frac{\vec p}{p}\int dp_1\frac{dq}{q^2}
		\sqrt{\frac{p_1(p+q)(p_1-q)}{p}}\\
		&\times\frac{g(p,\omega)+g(p_1,\omega)-g(p+q,\omega)-g(p_1-q,\omega)}{(1+e^{-\beta'\nu_F(p+p\zeta_x)})(1+e^{-\beta'\nu_F(p_1+p_1\zeta_x)})(1+e^{\beta'\nu_F(p+q+(p+q)\zeta_x)})(1+e^{\beta'\nu_F(p_1-q-(p_1-q)\zeta_x)})},
	\end{aligned}
	\label{alphalog.eqn}
\end{equation} 
\end{widetext}
where in the collinear limit, we have used $p>0, p+q>0$ and $p_1-q>0$. Exchange terms are negligible when $\alpha\to 0$ where a logarithm dominates
and the Debye screening mass makes the above integral convergent in the $q\rightarrow 0$ limit~\cite{kashuba}.

As can be seen in Eq.~\eqref{alphalog.eqn}, the same-charge collision integral that is proportional to $\alpha^2 \ln{\alpha}$
becomes zero if the momentum-independent ansatz $g(p,\omega)\propto g(\omega)$ is employed. However, when this ansatz is used, 
there will be other non-collinear (and hence non-divergent) terms proportional to $\alpha^2$ (rather than $\alpha^2\log\alpha$). 
These types of terms in the large logarithm limit contribute smaller eigenvalues and hence by Eq.~\eqref{spectral.eqn} dominate the distribution function. 

Hence up to logarithm corrections we are led to adopt the following ansatz:
\begin{equation}
\hat g(p,\omega)=v_F\beta^{'2} g(\omega),
\label{perturbation}	
\end{equation}     
where the coefficients are chosen so as to make $g(\omega)$ dimensionless. 
Finally, we integrate the functional ${\cal Q}[g]$ using the coordinate parametrization~\eqref{ellipse} to obtain
\begin{equation}
	\begin{aligned}	
		{\cal Q}[g]
		=\gamma\beta'\frac{\ln2}{4\pi}\Big[\gamma^2\kappa(\zeta)\alpha^2g(\omega)^2\! -\! i\omega \beta'g(\omega)^2\!-\!2g(\omega)\Big].
	\end{aligned}
\end{equation}
It should be noted that the parameter $\kappa$ depends on the absolute value of the tilt parameter and not its direction with respect to the electric field which is 
denoted by $\theta_t$. We can show that this fact by reparametrization of theta angle $\theta_l$ by $\theta_l-\theta_t$ if we write the measure $d^2l$ as $ldld\theta_l$,
where $l=k, k_1$ or $q$.
The function $\kappa(\zeta)$ has been numerically computed for different values of tilted parameters as follows (see appendix~\ref{AppendixC}):
\begin{center}
	\begin{tabular}{|c|c|c|c|c|c|}
	\hline
	$\zeta$	 &0	&0.2	&0.4	&0.6	&0.8\\
	\hline
	$\kappa$ &3.65	&3.83	&4.26	&5.37	&7.97\\
	\hline
	\end{tabular}
\end{center}

\begin{figure}[b]
	\includegraphics[width=1\linewidth]{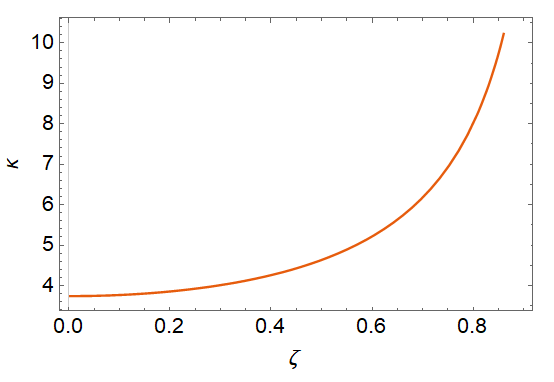}
	\caption{Numerically computed values of $\kappa$ as a function of $\zeta$. }
	\label{Fig1}
\end{figure}
We have plotted the results in Fig.~\ref{Fig1}. The Drude pole of conductivity coefficients Eq.~\eqref{sigmaxx} is given by
 \begin{equation}
	\begin{aligned}	
		\omega=\frac{\kappa(\zeta)\alpha^2k_BT}{i\hbar\sqrt{1-\zeta^2}},
	\end{aligned}
	\label{broadening.eqn}
\end{equation}
which shows that the frequency of absorption enhanced than graphene (i.e. $\zeta=0$ Dirac fermions) by factor of the form $\propto(1-\zeta^2)^{-\frac{1}{2}}$.
This can be regarded as a direct manifestation of the underlying spacetime metric~\eqref{metricds.eqn}. 
\begin{figure}[t]
	\includegraphics[width=1\linewidth]{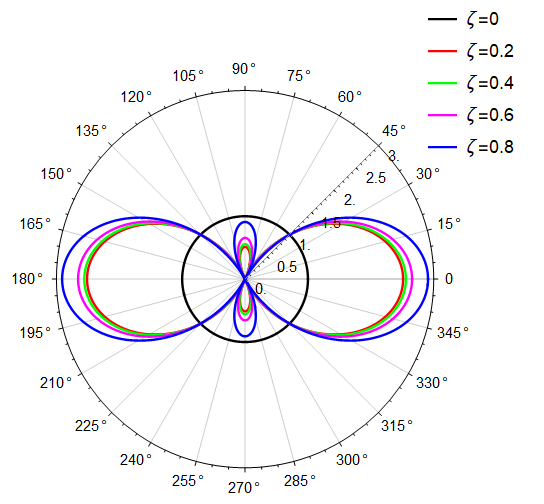}
	\caption{Angular dependence of the dimensionless Drude pole strength $\chi$. }
	\label{Fig2}
\end{figure}

Finally, we extremize the functional ${\cal Q}[g]$ with respect to the $g(\omega)$ to find the perturbation ansatz ~\eqref{perturbation} as 
 \begin{equation}
 	\hat g(p,\omega)=\frac{\frac{v_F}{T}}{-i\omega \sqrt{1-\zeta^2}+{\kappa}(\zeta)\alpha^2T}.
 \end{equation}
Then, the intraband conductivity coefficients can be read off as follows
 \begin{equation}\label{sigmaxx}
	\begin{aligned}	
		\sigma_{\rm intra}^{xx}&=\frac{e^2}{h}\frac{2Nk_BT \ln2(1-\zeta^2)}{(-i\hbar\omega\sqrt{1-\zeta^2}+\kappa(\zeta)\alpha^2k_BT)\zeta^2}\times\\&\qquad\times\bigg[-\frac{\zeta^2\zeta_x^2-\zeta_y^2}{(1-\zeta^2)^{\frac{3}{2}}}+\frac{1}{\zeta^2}\Big(1+\frac{2\zeta^2-1}{(1-\zeta^2)^{\frac{3}{2}}}\Big)(\zeta_x^2-\zeta_y^2)\bigg],
	\end{aligned}
\end{equation}
\begin{equation}
	\begin{aligned}	
		\sigma_{\rm intra}^{xy}&=\frac{e^2}{h}\frac{2Nk_BT \ln2(1-\zeta^2)}{(-i\hbar\omega\sqrt{1-\zeta^2}+\kappa(\zeta)\alpha^2k_BT)\zeta^2}\times \\&\qquad\times\bigg[\frac{2}{\zeta^2}\Big(1+\frac{2\zeta^2-1}{(1-\zeta^2)^{\frac{3}{2}}}\Big)-\frac{1+\zeta^2}{(1-\zeta^2)^{\frac{3}{2}}}\bigg]\zeta_x\zeta_y.
	\end{aligned}
\end{equation}
The above equations indicate that, the tilting of the Dirac cone increases the broadening Drude frequency by $\kappa(\zeta)\times\frac{1}{\sqrt{1-\zeta^2}}$. 
In addition to that, the strength of the Drude pole is also enhanced. 
To quantify the effect of the tilt, let us define a dimensionless conductivity by
	\bea
	 \chi=\frac{\mbox{Res}(\sigma^{xx}_{\zeta})}{\mbox{Res}(\sigma^{xx}_{\zeta=0})},
	 \label{dimless.eqn}
	 \eea
where the Res means the residue at the Drude pole and defiens the strength of the Drude pole. In Fig.~\ref{Fig2}
the polar plot of $\chi$ is shown. 
As is evident, the strength of the pole are enhanced 2$\sim$3 times for different values of $\zeta$ when angle between $E_x$ and $\vec\zeta$ are not too large.

\section{Summary and outlook}
{\label{discuss.sec}}
When electrons are in a spacetime with non-zero $\zeta$, the time intervals measured in parts of spacetime that have
zero $\zeta$ are related to those in the spacetime~\eqref{metricds.eqn} by
\be
   dt=\frac{dt_0}{\sqrt{1-\zeta^2}}.
\ee
This means that the corresponding frequencies are related by 
\be
   \delta\omega_0=\frac{\delta\omega}{\sqrt{1-\zeta^2}}.
\ee
Therefore, the part of the broadening in Eq.~\eqref{broadening.eqn} that is proportional to $(1-\zeta^2)^{-1/2}$,
is a manifestation of an underlying spacetime structure given in Eq.~\eqref{metricds.eqn}.
The additional broadening
contained in $\kappa(\zeta)$ (see Fig.~\ref{Fig1}) arise from particular form of the Coulomb
integrals and can be attributed to the role of interaction in a background spacetime with non-zero tilt parameter $\zeta$~\cite{SaharCovariance}.
In addition to broadening of the Drude peak, the strength of the pole is also enhanced as given by Eq.~\eqref{dimless.eqn} and plotted in Fig.~\ref{Fig2}. 
It is important to note that in this calculation we have not explicitly used the metric Eq.~\eqref{metricds.eqn}, nevertheless, the 
above redshift factors appear quite naturally, indicating that the optical process and the interactions
are taking place in an underlying metric structure.The effects encoded in $\kappa(\zeta)$ can not be captured by 
a simple coordinate transformation. This is because the Coulomb forces (being a force between electrons in a material) 
can not be covariantly expressed in the above spacetime, as the Coulomb forces in a 2D material are mediated by 
photons that mostly propagate in 3D (Minkowski) spacetime outside the material.

Therefore the conductivity of tilted Dirac-cone fermions contains a wealth of information about the tilt parameter $\zeta$. 
This parameter defines an underlying spacetime metric~\eqref{metricds.eqn}. If for a tilted Dirac cone system, one can enforce the tilt
parameter $\zeta$ to vary in space over the length scales $\gtrsim\lambda$, where $\lambda$ is the wavelength of the light, 
from the local measurement of the optical conductivity and only the width and strength of the Drude peak, one can reconstruct 
the $\zeta(\vec x)$ from which the full structure of the emergent spacetime can be determined.

\section{acknowledgments}
MT and SAJ. appreciates research deputy of Sharif University of Technology, Grant No. G960214 and Iran Science Elites Foundation (ISEF). AM thanks Jiayue Yang for identifying an error in the Fig.~\ref{Fig1}.  

\appendix{}
\section{}
\label{AppendixA}

The equilibrium distribution function in the covariant form is $f^0_s(p,E_s)=\frac{1}{1+e^{\beta U^{\mu}p_{\mu}}}$, where $U^{\mu}$ is the normalized fluid velocity ($U^{\mu}U_{\mu}=-1$) with respect to the metric~\eqref{metricds.eqn} and $p_{\mu}=(E_s,\vec p)$ is the energy-momentum four vector.

 For $+$ charge particles, we have the following amplitude functions:
\begin{equation}
	\begin{aligned}	
	&R_1(p,p_1,q)\\
	&=4\left[ T_{+--+}(p,p_1,q)-T_{+-+-}(p,p_1,-p-q+p_1)|^2\right.\\
	&\left.+3| T_{+--+}(p,p_1,q)|^2+3| T_{+-+-}(p,p_1,-p-q+p_1)|^2\right],
	\end{aligned}
\end{equation}
and
\begin{equation}
	\begin{aligned}	
		&R_2(p,p_1,q)\\
		&=4\left[\frac{1}{2}| ( T_{++++}(p,p_1,q)-T_{++++}(p,p_1,p_1-q-p))|^2\right.\\
		&\left.+3| T_{++++}(p,p_1,q)|^2+3| T_{++++}(p,p_1,p_1-q-p)|^2\right].
	\end{aligned}
\end{equation} 

 \section{}
  \label{AppendixB}
 In this appendix, we look at the Boltzmann equation from variational principle in more detail. 
 Our discussion in this section is based on the classic text of Ziman~\cite{ziman2001electrons}. 
 We consider elastic scattering of a particle from a state with momentum $p$ to state $p'$. The probability of this scattering is given by:
 \begin{equation}
 	\mathcal{P}^{p'}_{p}=f_p(1-f_{p'})\mathcal{D}^{p'}_{p}dp',
 \end{equation} 
where $\mathcal{D}^{p'}_{p}$ is the intrinsic transition amplitude from state $p$ to $p'$ which can be computed from Feynman diagrams. For time reversal interactions, we have following condition:
 \begin{equation}
 	\mathcal{D}^{p'}_{p}=\mathcal{D}^{p}_{p'}. 
 \end{equation} 
The scattering part of this interaction can be summarized as follows:
\begin{equation}
	\sim \int \Big(f_p(1-f_{p'})-f_{p'}(1-f_p)\Big)\mathcal{D}^{p'}_{p}dp'.
\end{equation} 
The linearized Collision part around the equilibrium distribution function $f^{0}_p$ is given by:
\begin{equation}
	\sim \int \Big((f_{p'}-f^{0}_{p'})-(f_p-f^{0}_{p'})\Big)\mathcal{D}^{p'}_{p}dp'.
\end{equation}
We assume that non-equilibrium distribution function is caused by turning on an electric field and a temperature gradient, so Boltzmann equation reads:
\begin{equation}
  \begin{aligned}
&-v_p.\frac{\partial f^{0}_p}{\partial T}\nabla T-v_p.e\frac{\partial f^{0}_p}{\partial\epsilon_p}\vec E\\
&=\int \Big((f_{p'}-f^{0}_{p'})-(f_p-f^{0}_{p'})\Big)\mathcal{D}^{p'}_{p}dp',
  \end{aligned}
\end{equation}
where $\epsilon_p$ is the energy of the particles. It is convenient to write non-equilibrium distribution function as follows:
\begin{equation}
	\begin{aligned}
	f_p\equiv f^{0}_{p}-\Phi_p\frac{\partial f^{0}_{p}}{\partial \epsilon_{p}}.
	\label{variation}
	\end{aligned}
\end{equation} 
 So Boltzmann equation is transformed to the following form:
\begin{equation}
  \begin{aligned}
 	&-v_p.\frac{\partial f^{0}_{p}}{\partial T}\nabla T-v_p.e\frac{\partial f^{0}_{p}}{\partial\epsilon_{p}}\vec E\\
 	&=\frac{1}{k_BT}\int (\Phi_p-\Phi_p')f^{0}_{p'}(1-f^{0}_{p'})\mathcal{D}^{p'}_{p}dp'\\
 	&=\frac{1}{k_BT}\int (\Phi_p-\Phi_p')\mathcal{P}^{p'}_{p}dp'. 
 	\label{elastic}
 \end{aligned}
\end{equation}
We can generalized the above equation for collisions between particles in a time reversal interaction. For the scattering of the following form: 
\begin{equation}
	\begin{aligned}
		\vec p+\vec p'=\vec p''+\vec p'''. 
	\end{aligned}
\end{equation}
Similarly we have the intrinsic transition amplitude
\begin{equation}
	\begin{aligned}
		\mathcal{D}^{p''p'''}_{pp'}dp'dp''dp'''
	\end{aligned}
\end{equation}
that denotes the rate of scattering of particle with momentum $\vec p$ from a particle with momentum $\vec p'$ (in the range $dp'$) into the particles with momentum $p''$ and $p'''$. The collision part in the Boltzmann equation is proportional to
\begin{eqnarray}
	&&\int\left[f_pf_{p'}(1-f_{p''})(1-f_{p'''})-f_{p''}f_{p'''}(1-f_{p})(1-f_{p'})\right]\nonumber\\
	&&\times \mathcal{D}^{p''p'''}_{pp'}dp'dp''dp'''.
\end{eqnarray}
Time reversal invariance of the interactions imposes
\begin{equation}
	\begin{aligned}
		\mathcal{D}^{p''p'''}_{pp'}=	\mathcal{D}^{pp'}_{p''p'''}. 
	\end{aligned}
\end{equation}
The equilibrium distribution function satisfies the relation
\begin{equation}
		f^0_pf^0_{p'}(1-f^0_{p''})(1-f^0_{p'''})=f^0_{p''}f^0_{p'''}(1-f^0_{p})(1-f^0_{p'}),
\end{equation}
and using the ansatz~\eqref{variation} and expansion around the equilibrium distribution function $f^0_p$, collision term is simplified to
\begin{equation}
\begin{aligned}
&\sim\frac{1}{k_BT}\int\Big(\Phi_p+\Phi_{p'}-\Phi_{p''}-\Phi_{p'''}\Big)\times\\
&f^0_pf^0_{p'}(1-f^0_{p''})(1-f^0_{p'''})\mathcal{D}^{p''p'''}_{pp'}dp'dp''dp'''\\
&=\frac{1}{k_BT}\int\Big(\Phi_p+\Phi_{p'}-\Phi_{p''}-\Phi_{p'''}\Big)\times\\
&\mathcal{P}^{p''p'''}_{pp'}dp'dp''dp'''.
\end{aligned}
\end{equation}
We can now explain variational principle for Boltzmann Eq.~\eqref{elastic} which can be summarized as an integral equation
\begin{equation}
	\begin{aligned}
		X(p)=\int(\Phi_p-\Phi_p')\mathcal{P}(p,p')dp'. 
		\label{inteq}
	\end{aligned}
\end{equation}
$X(p)$ in the left hand side of Eq.~\eqref{inteq} is a known function of temperature gradient or electric field. 
From the abstract point of view, Eq.~\eqref{inteq} has the following form:
\begin{equation}
	\begin{aligned}
	X=P\Phi.
	\label{Boltz}
	\end{aligned}
\end{equation}
This motivates to define an inner product as follows:
 \begin{equation}
	\begin{aligned}	
		\bra{\Phi_1}\ket{\Phi_2}=\int dk \Phi_1(k)\Phi_2(k). 
	\end{aligned}
\end{equation}
With respect to this inner product, the scattering operator $\mathcal{P}(p,p')$ is Hermitian as follows:
\begin{equation}
	\begin{aligned}	
		&\bra{\Phi_1}\ket{\mathcal P\Phi_2}=\bra{\mathcal P\Phi_1}\ket{\Phi_2}\\
		&=\frac{1}{2}\int(\Phi^1_p-\Phi^1_{p'})\mathcal{P}(p,p')(\Phi^2_p-\Phi^2_{p'})dpdp'.
	\end{aligned}
\end{equation}
The most important property $\mathcal{P}$ is that it is positive definite 
\begin{equation}
	\begin{aligned}
	\bra{\Phi}\ket{\mathcal{P}\Phi}\geq 0,
	\label{positivedefinity}
	\end{aligned}
\end{equation}
for all $\Phi$.

Boltzmann Eq.~\eqref{Boltz} implies that 
\begin{equation}
	\begin{aligned}
		\bra{\Phi}\ket{\mathcal{P}\Phi}=\bra{\Phi}\ket{X}. 
		\label{variationalprinciple}
	\end{aligned}
\end{equation}
The variational principle states that, among all the functions $\Phi$ that satisfy Eq.~\eqref{variationalprinciple}, 
the solution of Boltzmann Eq.~\eqref{Boltz} is the one that corresponds to the maximum of Eq.~\eqref{positivedefinity}.
\section{}
\label{AppendixC}
In this section we clarify how $\kappa$ is numerically computed. The relevant part of $\mathcal{Q}(g)$ is as follows
\begin{widetext}
	\begin{equation}
		\begin{aligned}	
			&{\cal Q}(g)=\frac{2\pi}{8}\int\frac{{\rm d}^2p}{(2\pi)^2}\frac{{\rm d}^2p_1}{(2\pi)^2}\frac{{\rm d}^2q}{(2\pi)^2}
			\bigg[\frac{\delta(E_{s,\vec p}+E_{-s,\vec p_1}-E_{s,\vec p+\vec q}-E_{-s,\vec p_1-\vec q})R_1(\vec p,\vec p_1,\vec q)}{(1+e^{-\beta'E_{s,\vec p}})(1+e^{-\beta'E_{-s,\vec p_1}})(1+e^{\beta'E_{s,\vec p+\vec q}})(1+e^{\beta'E_{-s,\vec p_1-\vec q}})}\times
			\\&\qquad\qquad\qquad\qquad\qquad\qquad\times\frac{1}{v_F^2}\Big(\frac{\partial E_{s,\vec p}}{\partial\vec p}\hat g(\vec p,\omega)+\frac{\partial E_{-s,\vec p_1}}{\partial\vec p_1}\hat g(\vec p_1,\omega)-\frac{\partial E_{s,\vec p+\vec q}}{\partial(\vec p+\vec q)}\hat g(\vec p+\vec q,\omega)
			-\frac{\partial E_{-s,\vec p_1-\vec q}}{\partial(\vec p_1-\vec q)}\hat g(\vec p_1-\vec q,\omega)\Big)^2\\
			&\qquad\qquad\qquad\qquad\qquad\qquad\qquad+\frac{\delta(E_{s,\vec p}+E_{s,\vec p_1}-E_{s,\vec p+\vec q}-E_{s,\vec p_1-\vec q}))R_2(\vec p,\vec p_1,\vec q)}{(1+e^{-\beta'E_{s,\vec p}})(1+e^{-\beta'E_{s,\vec p_1}})(1+e^{\beta'E_{s,\vec p+\vec q}})(1+e^{\beta'E_{s,\vec p_1-\vec q}})}\times\\
			&\qquad\qquad\qquad\qquad\qquad\qquad\times\frac{1}{v_F^2}\Big((\frac{\partial E_{s,\vec p}}{\partial\vec p}\hat g(\vec p,\omega)+\frac{\partial E_{s,\vec p_1}}{\partial\vec p_1}\hat g(\vec p_1,\omega)-\frac{\partial E_{s,\vec p+\vec q}}{\partial(\vec p+\vec q)}\hat g(\vec p+\vec q,\omega)-\frac{\partial E_{s,\vec p_1-\vec q}}{\partial(\vec p_1-\vec q)}\hat g(\vec p_1-\vec q,\omega)\Big)^2\bigg],
		\end{aligned}
	\end{equation} 
\end{widetext}
which consists of two terms, opposite charge and same charge scattering, which respectively are given as follows
\begin{widetext}
	\begin{equation}
		\begin{aligned}	
			&{\cal I}_1(g)=\frac{2\pi}{8}\int\frac{{\rm d}^2p}{(2\pi)^2}\frac{{\rm d}^2p_1}{(2\pi)^2}\frac{{\rm d}^2q}{(2\pi)^2}
			\bigg[\frac{\delta(E_{s,\vec p}+E_{-s,\vec p_1}-E_{s,\vec p+\vec q}-E_{-s,\vec p_1-\vec q})R_1(\vec p,\vec p_1,\vec q)}{(1+e^{-\beta'E_{s,\vec p}})(1+e^{-\beta'E_{-s,\vec p_1}})(1+e^{\beta'E_{s,\vec p+\vec q}})(1+e^{\beta'E_{-s,\vec p_1-\vec q}})}\times
			\\&\qquad\qquad\qquad\qquad\qquad\qquad\times\frac{1}{v_F^2}\Big(\frac{\partial E_{s,\vec p}}{\partial\vec p}\hat g(\vec p,\omega)+\frac{\partial E_{-s,\vec p_1}}{\partial\vec p_1}\hat g(\vec p_1,\omega)-\frac{\partial E_{s,\vec p+\vec q}}{\partial(\vec p+\vec q)}\hat g(\vec p+\vec q,\omega)
			-\frac{\partial E_{-s,\vec p_1-\vec q}}{\partial(\vec p_1-\vec q)}\hat g(\vec p_1-\vec q,\omega)\Big)^2\Bigg],
					\end{aligned}
		\end{equation} 
	\end{widetext}

	\begin{widetext}
		\begin{equation}
			\begin{aligned}		
			&{\cal I}_2(g)=\frac{2\pi}{8}\int\frac{{\rm d}^2p}{(2\pi)^2}\frac{{\rm d}^2p_1}{(2\pi)^2}\frac{{\rm d}^2q}{(2\pi)^2}\frac{\delta(E_{s,\vec p}+E_{s,\vec p_1}-E_{s,\vec p+\vec q}-E_{s,\vec p_1-\vec q}))R_2(\vec p,\vec p_1,\vec q)}{(1+e^{-\beta'E_{s,\vec p}})(1+e^{-\beta'E_{s,\vec p_1}})(1+e^{\beta'E_{s,\vec p+\vec q}})(1+e^{\beta'E_{s,\vec p_1-\vec q}})}\times\\
			&\qquad\qquad\qquad\qquad\qquad\qquad\times\frac{1}{v_F^2}\Big((\frac{\partial E_{s,\vec p}}{\partial\vec p}\hat g(\vec p,\omega)+\frac{\partial E_{s,\vec p_1}}{\partial\vec p_1}\hat g(\vec p_1,\omega)-\frac{\partial E_{s,\vec p+\vec q}}{\partial(\vec p+\vec q)}\hat g(\vec p+\vec q,\omega)-\frac{\partial E_{s,\vec p_1-\vec q}}{\partial(\vec p_1-\vec q)}\hat g(\vec p_1-\vec q,\omega)\Big)^2\bigg].
		\end{aligned}
	\end{equation} 
\end{widetext}
We first compute on ${\cal I}_2(g)$ which is simpler than ${\cal I}_1(g)$.

\subsection{Like charge particle scattering} By the following transformations $\vec k=\beta'v_F\vec p$,~$\vec k_1=\beta'v_F\vec p_1$ and $\vec q'=\beta'v_F\vec q$, the like charge particle scattering term ${\cal I}_2$ is simplified as follows
\bea
\label{I_2equation}
{\cal I}_2(g)=\frac{\beta'}{64\pi^3}\alpha^2g^2(\omega){\cal I}'_2(g)
\eea
where we have used explicit form of energy-momentum dispersion relation $E_{s,\vec p}=sv_Fp+v_F\vec p.\zeta$, $\hat g=v_F\beta^{'2}g(\omega)$,~$\alpha=\frac{e^2}{v_F\epsilon}$ and ${\cal I}'_2$ is given by 
	\begin{widetext}
	\begin{equation}
		\begin{aligned}		
			&{\cal I}'_2(g)=\int{\rm d}^2k{\rm d}^2k_1{\rm d}^2q'\frac{\delta(k+k_1-\mid k+q'\mid-\mid k_1-q'\mid)R'_2(\vec k,\vec k_1,\vec q')}{(1+e^{-k+\vec k.\vec\zeta})(1+e^{-k_1+\vec k_1.\vec\zeta})(1+e^{\mid k+q'\mid+(\vec k+\vec q').\vec\zeta})(1+e^{\mid k_1-q'\mid+(\vec k_1-\vec q').\vec\zeta})}\\
			&\times\Big(\frac{\vec k}{k}+\frac{\vec k_1}{k_1}-\frac{\vec k+\vec q}{\mid k+q\mid}-\frac{\vec k_1-\vec q}{\mid k_1-q\mid}\Big)^2
		\end{aligned}
	\end{equation} 
\end{widetext}
where the difference between $R'(\vec k,\vec k_1,\vec q')$ and $R'(\vec p,\vec p_1,\vec q)$ is that in the former $V(q')$ is $V(q')=\frac{1}{q'}$.

Using the parametrization~\eqref{ellipse}, the Dirac delta function can be eliminated as follows 
	\begin{widetext}
\bea
{\cal I}_2'(g)&=&\int {\rm d}^2k{\rm d}^2k_1\int {\rm d}^2q'\delta(k+k_1-\mid k+q'\mid-\mid k_1-q'\mid)...\nn\\
&=&\int {\rm d}^2k{\rm d}^2k_1\int_0^{2\pi}{\rm d}\theta\int_0^{\infty}{\rm d}\mu\frac{\mid k+k_1\mid^2}{4}(cosh(\mu)^2-cos(\theta)^2)\delta(k+k_1-cosh(\mu)\mid k+k_1\mid)...\nn\\
&=&\int {\rm d}^2k{\rm d}^2k_1\int_0^{2\pi}{\rm d}\theta\frac{\mid k+k_1\mid(cosh(\mu)^2-cos(\theta)^2)}{4sinh(\mu)}...
\eea 
\end{widetext}
where ... are other functions in the integral. Finally ${\cal I}_2(g)$ part is summarized as follows
\bea
{\cal I}_2(g)=\frac{\beta'}{4\times 64\pi^3}\alpha^2g^2(\omega){\cal I}^{"}_2
\eea
\bea
{\cal I}^{''}_2(g)&=&\int {\rm d}^2k{\rm d}^2k_1\int_0^{2\pi}\frac{\mid k+k_1\mid(cosh(\mu)^2-cos(\theta)^2)}{sinh(\mu)}...\nn\\
\eea
\subsection{Opposite charge particle scattering}
The rest of the computation is the same as like charge particle scattering besides the fact that the argument of delta function is a hyperbola
rather than an elliptic curve. We have the same relation for ${\cal I}_1(g)$ simillar to eq.~\eqref{I_2equation} as follows :

\bea
{\cal I}_1(g)=\frac{\beta'}{64\pi^3}\alpha^2g^2(\omega){\cal I}'_1(g)
\eea

and ${\cal I}'_1(g)$ simillarly is given by

	\begin{widetext}
	\bea
	{\cal I}_1'(g)&=&\int {\rm d}^2k{\rm d}^2k_1\int {\rm d}^2q'\delta(k-k_1-\mid k+q'\mid+\mid k_1-q'\mid)...\nn\\
	&=&\int {\rm d}^2k{\rm d}^2k_1\int_0^{2\pi}{\rm d}\theta\int_0^{\infty}{\rm d}\mu\frac{\mid k+k_1\mid^2}{4}(cosh(\mu)^2-cos(\theta)^2)\delta(k-k_1-cos(\theta)\mid k+k_1\mid)...\nn\\
	&=&2\int {\rm d}^2k{\rm d}^2k_1\int_0^{\infty}{\rm d}\mu\frac{\mid k+k_1\mid(cosh(\mu)^2-cos(\theta)^2)}{4sin(\theta)}...,
	\eea 
\end{widetext}
the argument of the Dirac delta function have two solution for $cos(\theta)=\frac{k-k_1}{\mid k+k_1\mid}$ in the range $0<\theta<2\pi$ and prefactor 2 is the origin of it. To be consistent with ${\cal I}^{''}_2$ part, we define ${\cal I}^{''}_1(g)$ as ${\cal I}^{''}_1(g)=4{\cal I}^'_1(g)$. The total contribution to ${\cal I}(g)$ is given by
\bea
{\cal I}(g)&=&{\cal I}_1(g)+{\cal I}_2(g)=(\frac{\gamma^3\beta'log 2}{4\pi})(\frac{{\cal I}^{''}_1(g)+{\cal I}^{''}_2(g)}{64\pi^2log 2\gamma^3})\alpha^2g(\omega)^2,\nn\\
&\equiv&(\frac{\gamma^3\beta'log 2}{4\pi})\kappa\alpha^2g^2(\omega).\eea

The integrals ${\cal I}^{''}_1(g)$ and ${\cal I}^{''}_1(g)$ are convergent and can be integrated numerically by mathematica.

\bibliography{mybib} 

\begin{thebibliography}{37}%
\makeatletter
\providecommand \@ifxundefined [1]{%
 \@ifx{#1\undefined}
}%
\providecommand \@ifnum [1]{%
 \ifnum #1\expandafter \@firstoftwo
 \else \expandafter \@secondoftwo
 \fi
}%
\providecommand \@ifx [1]{%
 \ifx #1\expandafter \@firstoftwo
 \else \expandafter \@secondoftwo
 \fi
}%
\providecommand \natexlab [1]{#1}%
\providecommand \enquote  [1]{``#1''}%
\providecommand \bibnamefont  [1]{#1}%
\providecommand \bibfnamefont [1]{#1}%
\providecommand \citenamefont [1]{#1}%
\providecommand \href@noop [0]{\@secondoftwo}%
\providecommand \href [0]{\begingroup \@sanitize@url \@href}%
\providecommand \@href[1]{\@@startlink{#1}\@@href}%
\providecommand \@@href[1]{\endgroup#1\@@endlink}%
\providecommand \@sanitize@url [0]{\catcode `\\12\catcode `\$12\catcode
  `\&12\catcode `\#12\catcode `\^12\catcode `\_12\catcode `\%12\relax}%
\providecommand \@@startlink[1]{}%
\providecommand \@@endlink[0]{}%
\providecommand \url  [0]{\begingroup\@sanitize@url \@url }%
\providecommand \@url [1]{\endgroup\@href {#1}{\urlprefix }}%
\providecommand \urlprefix  [0]{URL }%
\providecommand \Eprint [0]{\href }%
\providecommand \doibase [0]{http://dx.doi.org/}%
\providecommand \selectlanguage [0]{\@gobble}%
\providecommand \bibinfo  [0]{\@secondoftwo}%
\providecommand \bibfield  [0]{\@secondoftwo}%
\providecommand \translation [1]{[#1]}%
\providecommand \BibitemOpen [0]{}%
\providecommand \bibitemStop [0]{}%
\providecommand \bibitemNoStop [0]{.\EOS\space}%
\providecommand \EOS [0]{\spacefactor3000\relax}%
\providecommand \BibitemShut  [1]{\csname bibitem#1\endcsname}%
\let\auto@bib@innerbib\@empty
\bibitem [{\citenamefont {Haug}\ and\ \citenamefont {Jauho}(2008)}]{Jauho}%
  \BibitemOpen
  \bibfield  {author} {\bibinfo {author} {\bibfnamefont {H.}~\bibnamefont
  {Haug}}\ and\ \bibinfo {author} {\bibfnamefont {A.-P.}\ \bibnamefont
  {Jauho}},\ }\href {\doibase 10.1007/978-3-540-73564-9} {\emph {\bibinfo
  {title} {Quantum Kinetics in Transport and Optics of Semiconductors}}}\
  (\bibinfo  {publisher} {Springer Berlin Heidelberg},\ \bibinfo {year}
  {2008})\BibitemShut {NoStop}%
\bibitem [{\citenamefont {Ziman}(2001)}]{ziman2001electrons}%
  \BibitemOpen
  \bibfield  {author} {\bibinfo {author} {\bibfnamefont {J.~M.}\ \bibnamefont
  {Ziman}},\ }\href@noop {} {\emph {\bibinfo {title} {Electrons and phonons:
  the theory of transport phenomena in solids}}}\ (\bibinfo  {publisher}
  {Oxford university press},\ \bibinfo {year} {2001})\BibitemShut {NoStop}%
\bibitem [{\citenamefont {Girvin}\ and\ \citenamefont
  {Yang}(2019)}]{Girvin2019}%
  \BibitemOpen
  \bibfield  {author} {\bibinfo {author} {\bibfnamefont {S.~M.}\ \bibnamefont
  {Girvin}}\ and\ \bibinfo {author} {\bibfnamefont {K.}~\bibnamefont {Yang}},\
  }\href@noop {} {\emph {\bibinfo {title} {Modern Condensed Matter Physics}}}\
  (\bibinfo  {publisher} {Cambridge University Press},\ \bibinfo {address} {New
  York},\ \bibinfo {year} {2019})\BibitemShut {NoStop}%
\bibitem [{\citenamefont {Cercignani}\ and\ \citenamefont
  {Kremer}(2002)}]{Cercignani2002}%
  \BibitemOpen
  \bibfield  {author} {\bibinfo {author} {\bibfnamefont {C.}~\bibnamefont
  {Cercignani}}\ and\ \bibinfo {author} {\bibfnamefont {G.~M.}\ \bibnamefont
  {Kremer}},\ }\href {\doibase 10.1007/978-3-0348-8165-4} {\emph {\bibinfo
  {title} {The Relativistic Boltzmann Equation: Theory and Applications}}}\
  (\bibinfo  {publisher} {Birkhäuser Basel},\ \bibinfo {year}
  {2002})\BibitemShut {NoStop}%
\bibitem [{\citenamefont {Das~Sarma}\ \emph {et~al.}(2011)\citenamefont
  {Das~Sarma}, \citenamefont {Adam}, \citenamefont {Hwang},\ and\ \citenamefont
  {Rossi}}]{SarmaReview}%
  \BibitemOpen
  \bibfield  {author} {\bibinfo {author} {\bibfnamefont {S.}~\bibnamefont
  {Das~Sarma}}, \bibinfo {author} {\bibfnamefont {S.}~\bibnamefont {Adam}},
  \bibinfo {author} {\bibfnamefont {E.~H.}\ \bibnamefont {Hwang}}, \ and\
  \bibinfo {author} {\bibfnamefont {E.}~\bibnamefont {Rossi}},\ }\href
  {\doibase 10.1103/RevModPhys.83.407} {\bibfield  {journal} {\bibinfo
  {journal} {Rev. Mod. Phys.}\ }\textbf {\bibinfo {volume} {83}},\ \bibinfo
  {pages} {407} (\bibinfo {year} {2011})}\BibitemShut {NoStop}%
\bibitem [{\citenamefont {Xiao}\ \emph {et~al.}(2010)\citenamefont {Xiao},
  \citenamefont {Chang},\ and\ \citenamefont {Niu}}]{Niu2010}%
  \BibitemOpen
  \bibfield  {author} {\bibinfo {author} {\bibfnamefont {D.}~\bibnamefont
  {Xiao}}, \bibinfo {author} {\bibfnamefont {M.-C.}\ \bibnamefont {Chang}}, \
  and\ \bibinfo {author} {\bibfnamefont {Q.}~\bibnamefont {Niu}},\ }\href
  {\doibase 10.1103/RevModPhys.82.1959} {\bibfield  {journal} {\bibinfo
  {journal} {Rev. Mod. Phys.}\ }\textbf {\bibinfo {volume} {82}},\ \bibinfo
  {pages} {1959} (\bibinfo {year} {2010})}\BibitemShut {NoStop}%
\bibitem [{\citenamefont {Stephanov}\ and\ \citenamefont
  {Yin}(2012)}]{Stephanov2012}%
  \BibitemOpen
  \bibfield  {author} {\bibinfo {author} {\bibfnamefont {M.~A.}\ \bibnamefont
  {Stephanov}}\ and\ \bibinfo {author} {\bibfnamefont {Y.}~\bibnamefont
  {Yin}},\ }\href {\doibase 10.1103/PhysRevLett.109.162001} {\bibfield
  {journal} {\bibinfo  {journal} {Phys. Rev. Lett.}\ }\textbf {\bibinfo
  {volume} {109}},\ \bibinfo {pages} {162001} (\bibinfo {year}
  {2012})}\BibitemShut {NoStop}%
\bibitem [{\citenamefont {Son}\ and\ \citenamefont
  {Spivak}(2013)}]{SonSpivak2013}%
  \BibitemOpen
  \bibfield  {author} {\bibinfo {author} {\bibfnamefont {D.~T.}\ \bibnamefont
  {Son}}\ and\ \bibinfo {author} {\bibfnamefont {B.~Z.}\ \bibnamefont
  {Spivak}},\ }\href {\doibase 10.1103/PhysRevB.88.104412} {\bibfield
  {journal} {\bibinfo  {journal} {Phys. Rev. B}\ }\textbf {\bibinfo {volume}
  {88}},\ \bibinfo {pages} {104412} (\bibinfo {year} {2013})}\BibitemShut
  {NoStop}%
\bibitem [{\citenamefont {Kashuba}(2008)}]{kashuba}%
  \BibitemOpen
  \bibfield  {author} {\bibinfo {author} {\bibfnamefont {A.~B.}\ \bibnamefont
  {Kashuba}},\ }\href {\doibase 10.1103/PhysRevB.78.085415} {\bibfield
  {journal} {\bibinfo  {journal} {Phys. Rev. B}\ }\textbf {\bibinfo {volume}
  {78}},\ \bibinfo {pages} {085415} (\bibinfo {year} {2008})}\BibitemShut
  {NoStop}%
\bibitem [{\citenamefont {Sachdev}(2007)}]{SachdevBook}%
  \BibitemOpen
  \bibfield  {author} {\bibinfo {author} {\bibfnamefont {S.}~\bibnamefont
  {Sachdev}},\ }\href@noop {} {\emph {\bibinfo {title} {Quantum phase
  transitions}}}\ (\bibinfo  {publisher} {Wiley Online Library},\ \bibinfo
  {year} {2007})\BibitemShut {NoStop}%
\bibitem [{\citenamefont {Fritz}\ \emph {et~al.}(2008)\citenamefont {Fritz},
  \citenamefont {Schmalian}, \citenamefont {M{\"u}ller},\ and\ \citenamefont
  {Sachdev}}]{sachdev08}%
  \BibitemOpen
  \bibfield  {author} {\bibinfo {author} {\bibfnamefont {L.}~\bibnamefont
  {Fritz}}, \bibinfo {author} {\bibfnamefont {J.}~\bibnamefont {Schmalian}},
  \bibinfo {author} {\bibfnamefont {M.}~\bibnamefont {M{\"u}ller}}, \ and\
  \bibinfo {author} {\bibfnamefont {S.}~\bibnamefont {Sachdev}},\ }\href
  {https://journals.aps.org/prb/abstract/10.1103/PhysRevB.78.085416} {\bibfield
   {journal} {\bibinfo  {journal} {Phys. Rev. B}\ }\textbf {\bibinfo {volume}
  {78}},\ \bibinfo {pages} {085416} (\bibinfo {year} {2008})}\BibitemShut
  {NoStop}%
\bibitem [{\citenamefont {M\"uller}\ \emph {et~al.}(2009)\citenamefont
  {M\"uller}, \citenamefont {Schmalian},\ and\ \citenamefont
  {Fritz}}]{muller2009graphene}%
  \BibitemOpen
  \bibfield  {author} {\bibinfo {author} {\bibfnamefont {M.}~\bibnamefont
  {M\"uller}}, \bibinfo {author} {\bibfnamefont {J.}~\bibnamefont {Schmalian}},
  \ and\ \bibinfo {author} {\bibfnamefont {L.}~\bibnamefont {Fritz}},\ }\href
  {\doibase 10.1103/PhysRevLett.103.025301} {\bibfield  {journal} {\bibinfo
  {journal} {Phys. Rev. Lett.}\ }\textbf {\bibinfo {volume} {103}},\ \bibinfo
  {pages} {025301} (\bibinfo {year} {2009})}\BibitemShut {NoStop}%
\bibitem [{\citenamefont {Lucas}\ and\ \citenamefont
  {Fong}(2018)}]{lucas2018hydrodynamics}%
  \BibitemOpen
  \bibfield  {author} {\bibinfo {author} {\bibfnamefont {A.}~\bibnamefont
  {Lucas}}\ and\ \bibinfo {author} {\bibfnamefont {K.~C.}\ \bibnamefont
  {Fong}},\ }\href {\doibase 10.1088/1361-648x/aaa274} {\bibfield  {journal}
  {\bibinfo  {journal} {Journal of Physics: Condensed Matter}\ }\textbf
  {\bibinfo {volume} {30}},\ \bibinfo {pages} {053001} (\bibinfo {year}
  {2018})}\BibitemShut {NoStop}%
\bibitem [{\citenamefont {Crossno}\ \emph {et~al.}(2016)\citenamefont
  {Crossno}, \citenamefont {Shi}, \citenamefont {Wang}, \citenamefont {Liu},
  \citenamefont {Harzheim}, \citenamefont {Lucas}, \citenamefont {Sachdev},
  \citenamefont {Kim}, \citenamefont {Taniguchi}, \citenamefont {Watanabe},
  \citenamefont {Ohki},\ and\ \citenamefont {Fong}}]{Crossno2016}%
  \BibitemOpen
  \bibfield  {author} {\bibinfo {author} {\bibfnamefont {J.}~\bibnamefont
  {Crossno}}, \bibinfo {author} {\bibfnamefont {J.~K.}\ \bibnamefont {Shi}},
  \bibinfo {author} {\bibfnamefont {K.}~\bibnamefont {Wang}}, \bibinfo {author}
  {\bibfnamefont {X.}~\bibnamefont {Liu}}, \bibinfo {author} {\bibfnamefont
  {A.}~\bibnamefont {Harzheim}}, \bibinfo {author} {\bibfnamefont
  {A.}~\bibnamefont {Lucas}}, \bibinfo {author} {\bibfnamefont
  {S.}~\bibnamefont {Sachdev}}, \bibinfo {author} {\bibfnamefont
  {P.}~\bibnamefont {Kim}}, \bibinfo {author} {\bibfnamefont {T.}~\bibnamefont
  {Taniguchi}}, \bibinfo {author} {\bibfnamefont {K.}~\bibnamefont {Watanabe}},
  \bibinfo {author} {\bibfnamefont {T.~A.}\ \bibnamefont {Ohki}}, \ and\
  \bibinfo {author} {\bibfnamefont {K.~C.}\ \bibnamefont {Fong}},\ }\href
  {\doibase 10.1126/science.aad0343} {\bibfield  {journal} {\bibinfo  {journal}
  {Science}\ }\textbf {\bibinfo {volume} {351}},\ \bibinfo {pages} {1058}
  (\bibinfo {year} {2016})}\BibitemShut {NoStop}%
\bibitem [{\citenamefont {Bandurin}\ \emph {et~al.}(2016)\citenamefont
  {Bandurin}, \citenamefont {Torre}, \citenamefont {Kumar}, \citenamefont
  {Shalom}, \citenamefont {Tomadin}, \citenamefont {Principi}, \citenamefont
  {Auton}, \citenamefont {Khestanova}, \citenamefont {Novoselov}, \citenamefont
  {Grigorieva}, \citenamefont {Ponomarenko}, \citenamefont {Geim},\ and\
  \citenamefont {Polini}}]{Bandurin2016}%
  \BibitemOpen
  \bibfield  {author} {\bibinfo {author} {\bibfnamefont {D.~A.}\ \bibnamefont
  {Bandurin}}, \bibinfo {author} {\bibfnamefont {I.}~\bibnamefont {Torre}},
  \bibinfo {author} {\bibfnamefont {R.~K.}\ \bibnamefont {Kumar}}, \bibinfo
  {author} {\bibfnamefont {M.~B.}\ \bibnamefont {Shalom}}, \bibinfo {author}
  {\bibfnamefont {A.}~\bibnamefont {Tomadin}}, \bibinfo {author} {\bibfnamefont
  {A.}~\bibnamefont {Principi}}, \bibinfo {author} {\bibfnamefont {G.~H.}\
  \bibnamefont {Auton}}, \bibinfo {author} {\bibfnamefont {E.}~\bibnamefont
  {Khestanova}}, \bibinfo {author} {\bibfnamefont {K.~S.}\ \bibnamefont
  {Novoselov}}, \bibinfo {author} {\bibfnamefont {I.~V.}\ \bibnamefont
  {Grigorieva}}, \bibinfo {author} {\bibfnamefont {L.~A.}\ \bibnamefont
  {Ponomarenko}}, \bibinfo {author} {\bibfnamefont {A.~K.}\ \bibnamefont
  {Geim}}, \ and\ \bibinfo {author} {\bibfnamefont {M.}~\bibnamefont
  {Polini}},\ }\href {\doibase 10.1126/science.aad0201} {\bibfield  {journal}
  {\bibinfo  {journal} {Science}\ }\textbf {\bibinfo {volume} {351}},\ \bibinfo
  {pages} {1055} (\bibinfo {year} {2016})}\BibitemShut {NoStop}%
\bibitem [{\citenamefont {Bandurin}\ \emph {et~al.}(2018)\citenamefont
  {Bandurin}, \citenamefont {Shytov}, \citenamefont {Levitov}, \citenamefont
  {Kumar}, \citenamefont {Berdyugin}, \citenamefont {Shalom}, \citenamefont
  {Grigorieva}, \citenamefont {Geim},\ and\ \citenamefont
  {Falkovich}}]{Bandurin2018}%
  \BibitemOpen
  \bibfield  {author} {\bibinfo {author} {\bibfnamefont {D.~A.}\ \bibnamefont
  {Bandurin}}, \bibinfo {author} {\bibfnamefont {A.~V.}\ \bibnamefont
  {Shytov}}, \bibinfo {author} {\bibfnamefont {L.~S.}\ \bibnamefont {Levitov}},
  \bibinfo {author} {\bibfnamefont {R.~K.}\ \bibnamefont {Kumar}}, \bibinfo
  {author} {\bibfnamefont {A.~I.}\ \bibnamefont {Berdyugin}}, \bibinfo {author}
  {\bibfnamefont {M.~B.}\ \bibnamefont {Shalom}}, \bibinfo {author}
  {\bibfnamefont {I.~V.}\ \bibnamefont {Grigorieva}}, \bibinfo {author}
  {\bibfnamefont {A.~K.}\ \bibnamefont {Geim}}, \ and\ \bibinfo {author}
  {\bibfnamefont {G.}~\bibnamefont {Falkovich}},\ }\href
  {https://doi.org/10.1038%2Fs41467-018-07004-4} {\bibfield  {journal}
  {\bibinfo  {journal} {Nature Communications}\ }\textbf {\bibinfo {volume}
  {9}} (\bibinfo {year} {2018})}\BibitemShut {NoStop}%
\bibitem [{\citenamefont {Zhou}\ \emph {et~al.}(2014)\citenamefont {Zhou},
  \citenamefont {Dong}, \citenamefont {Oganov}, \citenamefont {Zhu},
  \citenamefont {Tian},\ and\ \citenamefont {Wang}}]{Zhou2014}%
  \BibitemOpen
  \bibfield  {author} {\bibinfo {author} {\bibfnamefont {X.-F.}\ \bibnamefont
  {Zhou}}, \bibinfo {author} {\bibfnamefont {X.}~\bibnamefont {Dong}}, \bibinfo
  {author} {\bibfnamefont {A.~R.}\ \bibnamefont {Oganov}}, \bibinfo {author}
  {\bibfnamefont {Q.}~\bibnamefont {Zhu}}, \bibinfo {author} {\bibfnamefont
  {Y.}~\bibnamefont {Tian}}, \ and\ \bibinfo {author} {\bibfnamefont {H.-T.}\
  \bibnamefont {Wang}},\ }\href {\doibase 10.1103/PhysRevLett.112.085502}
  {\bibfield  {journal} {\bibinfo  {journal} {Phys. Rev. Lett.}\ }\textbf
  {\bibinfo {volume} {112}},\ \bibinfo {pages} {085502} (\bibinfo {year}
  {2014})}\BibitemShut {NoStop}%
\bibitem [{\citenamefont {Lopez-Bezanilla}\ and\ \citenamefont
  {Littlewood}(2016)}]{Lopez2016}%
  \BibitemOpen
  \bibfield  {author} {\bibinfo {author} {\bibfnamefont {A.}~\bibnamefont
  {Lopez-Bezanilla}}\ and\ \bibinfo {author} {\bibfnamefont {P.~B.}\
  \bibnamefont {Littlewood}},\ }\href {\doibase 10.1103/PhysRevB.93.241405}
  {\bibfield  {journal} {\bibinfo  {journal} {Phys. Rev. B}\ }\textbf {\bibinfo
  {volume} {93}},\ \bibinfo {pages} {241405} (\bibinfo {year}
  {2016})}\BibitemShut {NoStop}%
\bibitem [{\citenamefont {Tan}\ \emph {et~al.}(2021)\citenamefont {Tan},
  \citenamefont {Yan}, \citenamefont {Zhao}, \citenamefont {Guo},\ and\
  \citenamefont {Chang}}]{MoS2Tilted}%
  \BibitemOpen
  \bibfield  {author} {\bibinfo {author} {\bibfnamefont {C.-Y.}\ \bibnamefont
  {Tan}}, \bibinfo {author} {\bibfnamefont {C.-X.}\ \bibnamefont {Yan}},
  \bibinfo {author} {\bibfnamefont {Y.-H.}\ \bibnamefont {Zhao}}, \bibinfo
  {author} {\bibfnamefont {H.}~\bibnamefont {Guo}}, \ and\ \bibinfo {author}
  {\bibfnamefont {H.-R.}\ \bibnamefont {Chang}},\ }\href
  {https://link.aps.org/doi/10.1103/PhysRevB.103.125425} {\bibfield  {journal}
  {\bibinfo  {journal} {Phys. Rev. B}\ }\textbf {\bibinfo {volume} {103}},\
  \bibinfo {pages} {125425} (\bibinfo {year} {2021})}\BibitemShut {NoStop}%
\bibitem [{\citenamefont {Kajita}\ \emph {et~al.}(2014)\citenamefont {Kajita},
  \citenamefont {Nishio}, \citenamefont {Tajima}, \citenamefont {Suzumura},\
  and\ \citenamefont {Kobayashi}}]{Kajita2014Review}%
  \BibitemOpen
  \bibfield  {author} {\bibinfo {author} {\bibfnamefont {K.}~\bibnamefont
  {Kajita}}, \bibinfo {author} {\bibfnamefont {Y.}~\bibnamefont {Nishio}},
  \bibinfo {author} {\bibfnamefont {N.}~\bibnamefont {Tajima}}, \bibinfo
  {author} {\bibfnamefont {Y.}~\bibnamefont {Suzumura}}, \ and\ \bibinfo
  {author} {\bibfnamefont {A.}~\bibnamefont {Kobayashi}},\ }\href
  {https://doi.org/10.7566%2Fjpsj.83.072002} {\bibfield  {journal} {\bibinfo
  {journal} {Journal of the Physical Society of Japan}\ }\textbf {\bibinfo
  {volume} {83}},\ \bibinfo {pages} {072002} (\bibinfo {year}
  {2014})}\BibitemShut {NoStop}%
\bibitem [{\citenamefont {Katayama}\ \emph {et~al.}(2006)\citenamefont
  {Katayama}, \citenamefont {Kobayashi},\ and\ \citenamefont
  {Suzumura}}]{Katayama2006}%
  \BibitemOpen
  \bibfield  {author} {\bibinfo {author} {\bibfnamefont {S.}~\bibnamefont
  {Katayama}}, \bibinfo {author} {\bibfnamefont {A.}~\bibnamefont {Kobayashi}},
  \ and\ \bibinfo {author} {\bibfnamefont {Y.}~\bibnamefont {Suzumura}},\
  }\href {\doibase 10.1143/jpsj.75.054705} {\bibfield  {journal} {\bibinfo
  {journal} {Journal of the Physical Society of Japan}\ }\textbf {\bibinfo
  {volume} {75}},\ \bibinfo {pages} {054705} (\bibinfo {year}
  {2006})}\BibitemShut {NoStop}%
\bibitem [{\citenamefont {Suzumura}\ and\ \citenamefont
  {Kobayashi}(2012)}]{Suzumura2012}%
  \BibitemOpen
  \bibfield  {author} {\bibinfo {author} {\bibfnamefont {Y.}~\bibnamefont
  {Suzumura}}\ and\ \bibinfo {author} {\bibfnamefont {A.}~\bibnamefont
  {Kobayashi}},\ }\href {\doibase 10.3390/cryst2020266} {\bibfield  {journal}
  {\bibinfo  {journal} {Crystals}\ }\textbf {\bibinfo {volume} {2}},\ \bibinfo
  {pages} {266} (\bibinfo {year} {2012})}\BibitemShut {NoStop}%
\bibitem [{\citenamefont {Volovik}(2016)}]{Volovik2016}%
  \BibitemOpen
  \bibfield  {author} {\bibinfo {author} {\bibfnamefont {G.~E.}\ \bibnamefont
  {Volovik}},\ }\href {\doibase 10.1134/s0021364016210050} {\bibfield
  {journal} {\bibinfo  {journal} {{JETP} Letters}\ }\textbf {\bibinfo {volume}
  {104}},\ \bibinfo {pages} {645} (\bibinfo {year} {2016})}\BibitemShut
  {NoStop}%
\bibitem [{\citenamefont {Farajollahpour}\ \emph {et~al.}(2019)\citenamefont
  {Farajollahpour}, \citenamefont {Faraei},\ and\ \citenamefont
  {Jafari}}]{Tohid2019Spacetime}%
  \BibitemOpen
  \bibfield  {author} {\bibinfo {author} {\bibfnamefont {T.}~\bibnamefont
  {Farajollahpour}}, \bibinfo {author} {\bibfnamefont {Z.}~\bibnamefont
  {Faraei}}, \ and\ \bibinfo {author} {\bibfnamefont {S.~A.}\ \bibnamefont
  {Jafari}},\ }\href {\doibase 10.1103/PhysRevB.99.235150} {\bibfield
  {journal} {\bibinfo  {journal} {Phys. Rev. B}\ }\textbf {\bibinfo {volume}
  {99}},\ \bibinfo {pages} {235150} (\bibinfo {year} {2019})}\BibitemShut
  {NoStop}%
\bibitem [{\citenamefont {Liang}\ and\ \citenamefont
  {Ojanen}(2019)}]{Ojanen2019}%
  \BibitemOpen
  \bibfield  {author} {\bibinfo {author} {\bibfnamefont {L.}~\bibnamefont
  {Liang}}\ and\ \bibinfo {author} {\bibfnamefont {T.}~\bibnamefont {Ojanen}},\
  }\href {\doibase 10.1103/PhysRevResearch.1.032006} {\bibfield  {journal}
  {\bibinfo  {journal} {Phys. Rev. Res.}\ }\textbf {\bibinfo {volume} {1}},\
  \bibinfo {pages} {032006} (\bibinfo {year} {2019})}\BibitemShut {NoStop}%
\bibitem [{\citenamefont {Jalali-Mola}\ and\ \citenamefont
  {Jafari}(2019)}]{SaharCovariance}%
  \BibitemOpen
  \bibfield  {author} {\bibinfo {author} {\bibfnamefont {Z.}~\bibnamefont
  {Jalali-Mola}}\ and\ \bibinfo {author} {\bibfnamefont {S.~A.}\ \bibnamefont
  {Jafari}},\ }\href {\doibase 10.1103/PhysRevB.100.075113} {\bibfield
  {journal} {\bibinfo  {journal} {Phys. Rev. B}\ }\textbf {\bibinfo {volume}
  {100}},\ \bibinfo {pages} {075113} (\bibinfo {year} {2019})}\BibitemShut
  {NoStop}%
\bibitem [{\citenamefont {Jafari}(2019)}]{Jafari2019}%
  \BibitemOpen
  \bibfield  {author} {\bibinfo {author} {\bibfnamefont {S.~A.}\ \bibnamefont
  {Jafari}},\ }\href {https://doi.org/10.1103%2Fphysrevb.100.045144} {\bibfield
   {journal} {\bibinfo  {journal} {Phys. Rev. B}\ }\textbf {\bibinfo {volume}
  {100}} (\bibinfo {year} {2019})}\BibitemShut {NoStop}%
\bibitem [{\citenamefont {Ryder}(2009)}]{RyderGR}%
  \BibitemOpen
  \bibfield  {author} {\bibinfo {author} {\bibfnamefont {L.}~\bibnamefont
  {Ryder}},\ }\href
  {https://www.cambridge.org/core/books/introduction-to-general-relativity/2ED32305ED2BD3132CEBB0C1BE4A6C4D}
  {\emph {\bibinfo {title} {Introduction to General Relativity}}}\ (\bibinfo
  {publisher} {Cambridge University Press},\ \bibinfo {year}
  {2009})\BibitemShut {NoStop}%
\bibitem [{\citenamefont {Schutz}(2009)}]{SchutzGR}%
  \BibitemOpen
  \bibfield  {author} {\bibinfo {author} {\bibfnamefont {B.}~\bibnamefont
  {Schutz}},\ }\href
  {https://www.cambridge.org/core/books/first-course-in-general-relativity/3805425203DD91A7436EF6E5F2082263}
  {\emph {\bibinfo {title} {A First Course in General Relativity}}}\ (\bibinfo
  {publisher} {Cambridge University Press},\ \bibinfo {year}
  {2009})\BibitemShut {NoStop}%
\bibitem [{\citenamefont {Weinberg}(1972)}]{weinbergbook}%
  \BibitemOpen
  \bibfield  {author} {\bibinfo {author} {\bibfnamefont {S.}~\bibnamefont
  {Weinberg}},\ }\href@noop {} {\emph {\bibinfo {title} {Gravitation and
  cosmology: principles and applications of the general theory of
  relativity}}}\ (\bibinfo {year} {1972})\BibitemShut {NoStop}%
\bibitem [{\citenamefont {Liu}\ \emph {et~al.}(2019)\citenamefont {Liu},
  \citenamefont {Gao}, \citenamefont {Mameda},\ and\ \citenamefont
  {Huang}}]{KineticCurved}%
  \BibitemOpen
  \bibfield  {author} {\bibinfo {author} {\bibfnamefont {Y.-C.}\ \bibnamefont
  {Liu}}, \bibinfo {author} {\bibfnamefont {L.-L.}\ \bibnamefont {Gao}},
  \bibinfo {author} {\bibfnamefont {K.}~\bibnamefont {Mameda}}, \ and\ \bibinfo
  {author} {\bibfnamefont {X.-G.}\ \bibnamefont {Huang}},\ }\href
  {https://link.aps.org/doi/10.1103/PhysRevD.99.085014} {\bibfield  {journal}
  {\bibinfo  {journal} {Phys. Rev. D}\ }\textbf {\bibinfo {volume} {99}},\
  \bibinfo {pages} {085014} (\bibinfo {year} {2019})}\BibitemShut {NoStop}%
\bibitem [{\citenamefont {M\"uller}\ \emph {et~al.}(2008)\citenamefont
  {M\"uller}, \citenamefont {Fritz},\ and\ \citenamefont
  {Sachdev}}]{muller2008quantum}%
  \BibitemOpen
  \bibfield  {author} {\bibinfo {author} {\bibfnamefont {M.}~\bibnamefont
  {M\"uller}}, \bibinfo {author} {\bibfnamefont {L.}~\bibnamefont {Fritz}}, \
  and\ \bibinfo {author} {\bibfnamefont {S.}~\bibnamefont {Sachdev}},\ }\href
  {\doibase 10.1103/PhysRevB.78.115406} {\bibfield  {journal} {\bibinfo
  {journal} {Phys. Rev. B}\ }\textbf {\bibinfo {volume} {78}},\ \bibinfo
  {pages} {115406} (\bibinfo {year} {2008})}\BibitemShut {NoStop}%
\bibitem [{\citenamefont {Jalali-Mola}\ and\ \citenamefont
  {Jafari}(2021)}]{jalali2021tilt}%
  \BibitemOpen
  \bibfield  {author} {\bibinfo {author} {\bibfnamefont {Z.}~\bibnamefont
  {Jalali-Mola}}\ and\ \bibinfo {author} {\bibfnamefont {S.~A.}\ \bibnamefont
  {Jafari}},\ }\href
  {https://journals.aps.org/prb/abstract/10.1103/PhysRevB.104.085152}
  {\bibfield  {journal} {\bibinfo  {journal} {Phys. Rev. B}\ }\textbf {\bibinfo
  {volume} {104}},\ \bibinfo {pages} {085152} (\bibinfo {year}
  {2021})}\BibitemShut {NoStop}%
\bibitem [{\citenamefont {Auslender}\ and\ \citenamefont
  {Katsnelson}(2007)}]{katsnelson2007}%
  \BibitemOpen
  \bibfield  {author} {\bibinfo {author} {\bibfnamefont {M.}~\bibnamefont
  {Auslender}}\ and\ \bibinfo {author} {\bibfnamefont {M.~I.}\ \bibnamefont
  {Katsnelson}},\ }\href {\doibase 10.1103/PhysRevB.76.235425} {\bibfield
  {journal} {\bibinfo  {journal} {Phys. Rev. B}\ }\textbf {\bibinfo {volume}
  {76}},\ \bibinfo {pages} {235425} (\bibinfo {year} {2007})}\BibitemShut
  {NoStop}%
\bibitem [{\citenamefont {Pathria}\ and\ \citenamefont
  {Beale}(2011)}]{pathria}%
  \BibitemOpen
  \bibfield  {author} {\bibinfo {author} {\bibfnamefont {R.~K.}\ \bibnamefont
  {Pathria}}\ and\ \bibinfo {author} {\bibfnamefont {P.~D.}\ \bibnamefont
  {Beale}},\ }\href@noop {} {\enquote {\bibinfo {title} {Statistical
  mechanics},}\ } (\bibinfo {year} {2011})\BibitemShut {NoStop}%
\bibitem [{\citenamefont {Sachdev}(1998)}]{sachdev97}%
  \BibitemOpen
  \bibfield  {author} {\bibinfo {author} {\bibfnamefont {S.}~\bibnamefont
  {Sachdev}},\ }\href {\doibase 10.1103/PhysRevB.57.7157} {\bibfield  {journal}
  {\bibinfo  {journal} {Phys. Rev. B}\ }\textbf {\bibinfo {volume} {57}},\
  \bibinfo {pages} {7157} (\bibinfo {year} {1998})}\BibitemShut {NoStop}%
\bibitem [{\citenamefont {Arnold}\ \emph {et~al.}(2000)\citenamefont {Arnold},
  \citenamefont {Moore},\ and\ \citenamefont {Yaffe}}]{Arnold2000}%
  \BibitemOpen
  \bibfield  {author} {\bibinfo {author} {\bibfnamefont {P.}~\bibnamefont
  {Arnold}}, \bibinfo {author} {\bibfnamefont {G.~D.}\ \bibnamefont {Moore}}, \
  and\ \bibinfo {author} {\bibfnamefont {L.~G.}\ \bibnamefont {Yaffe}},\ }\href
  {https://doi.org/10.1088/1126-6708/2000/11/001} {\bibfield  {journal}
  {\bibinfo  {journal} {J. High Energ. Phys.}\ }\textbf {\bibinfo {volume}
  {2000}},\ \bibinfo {pages} {001} (\bibinfo {year} {2000})}\BibitemShut
  {NoStop}%
\end{thebibliography}%
\end{document}